\documentclass[11pt]{article}
\pdfoutput=1
\usepackage{jheppub}

\usepackage{booktabs}
\usepackage{hyperref}
\usepackage{graphicx,xcolor}
\usepackage{esvect}
\usepackage{url}
\usepackage{amsmath,amssymb,amsfonts,bbm,dsfont}
\usepackage{multirow}
\usepackage{makecell}
\usepackage{textcomp}
\usepackage{mathtools}
\usepackage{amsthm}
\usepackage{latexsym}
\usepackage{braket}
\usepackage{hyperref}
\usepackage[normalem]{ulem}
\usepackage{braket}

\definecolor{bottle_green}{RGB}{0,106,78}
\definecolor{celadon_green}{RGB}{47,132,124}
\definecolor{emerald}{RGB}{80,220,100}
\definecolor{jade}{RGB}{0,168,107}

\newcommand{\Tr}[2]{\mathrm{Tr}_{#1}\left[ {#2} \right]} 

\newcommand{\q}{\mathrm{q}}

\newcommand{\cqstate}{\varrho}

\newcommand{\beq}{\begin{equation}}
\newcommand{\eeq}{\end{equation}}

\newtheorem{assumption}{Assumption}

\title{The weak field limit of quantum matter back-reacting on classical spacetime}
\author[a]{Isaac Layton,}
\author[a]{Jonathan Oppenheim,}
\author[a]{Andrea Russo,}
\author[a,b]{Zachary Weller-Davies.}
\affiliation[a]{Department of Physics and Astronomy, University College London, London WC1E 6BT, UK}
\affiliation[b]{Perimeter Institute for Theoretical Physics, Waterloo, Ontario, Canada}
\emailAdd{andrearusso.physics@gmail.com}
\emailAdd{j.oppenheim@ucl.ac.uk}
\emailAdd{isaac.layton.20@ucl.ac.uk}
\emailAdd{zwellerdavies@gmail.com}

\abstract{Consistent coupling of quantum and classical degrees of freedom exists so long as there is both diffusion of the classical degrees of freedom and decoherence of the quantum system.
In this paper, we derive the Newtonian limit of such classical-quantum (CQ) theories of gravity. Our results are obtained both via the gauge fixing of the recently proposed path integral theory of CQ general relativity and via the CQ master equation approach. In each case, we find the same weak field dynamics. We find that the Newtonian potential diffuses by an amount lower bounded by the decoherence rate into mass eigenstates. We also present our results as an unraveled system of stochastic differential equations for the trajectory of the hybrid classical-quantum state and provide a series of kernels for constructing figures of merit, which can be used to rule out part of the parameter space of classical-quantum theories of gravity by experimentally testing it via the decoherence-diffusion trade-off. We compare and contrast the weak field limit to previous models of classical Newtonian gravity coupled to quantum systems. Here, we find that the Newtonian potential and quantum state change in lock-step, with the flow of time being stochastic.

}
\begin{document}
\maketitle
\flushbottom

\section{Introduction}\label{sec: introduction}

In the last century, the unification of three of the four fundamental forces into the Standard model has fuelled the conviction that physics can, and should, be united under the framework of quantum theory. This goal has been frustrated by the lack of success in constructing a complete theory of quantum gravity valid beyond the Planck scale, which, according to these premises, is a necessary step towards a complete unification of physics.
Regardless of the physics above the Planck scale, it is generally understood that any theory of quantum gravity should have a low energy limit described by a quantum theory of a spin-2 boson. 

While it is widely believed that gravity should be quantised just like the other forces, there are reasons to think that gravity is unlike the others. It alone can be thought of as the theory of a universal space-time geometry in which matter lives. It is possible that the space-time metric is not quantum in nature. 
There have been many recent proposals for testing this question, by measuring low-energy gravitational phenomena which cannot be reproduced classically. Currently, the most promising experiments include those which aim to detect gravitationally induced entanglement in table-top experiments via spin entanglement witnesses \cite{kafri2013noise,bose2017spin,PhysRevLett.119.240402,Marshman_2020,pedernales2021enhancing,carney2021testing,christodoulou2022locally,Danielson:2021egj}. There have also been proposals to measure quantum features of gravity not reproducible classically without resorting to entanglement generation directly \cite{lami2023testing, Howl:2020isj}. Though feasible, current estimates suggest that the technology required to perform the experiments is decades away.

An alternative approach is to construct consistent theories in which gravity is not a quantum force and then consider experimental tests of these. Early attempts to construct such a theory were based on the semi-classical Einstein's equations \cite{moller1962theories, rosenfeld1963quantization,Hu:2008rga}. However, if
one takes the semi-classical equations as fundamental, then it is well-known that they are not consistent, leading to violations of the standard principles of quantum theory and a break-down of either operational no-signalling, the Born rule, or composition of quantum systems under the tensor product~\cite{Gisin:1989sx, gisin1990weinberg,polchinski1991weinberg,eppley1977necessity,galley2021nogo}. Moreover, even if one prefers to consider classical gravity only as an effective theory, the semi-classical equations result in pathological behaviour when quantum fluctuations are large \cite{Ford:1982wu,Kuo:1993if,wald1977back,Hu:2008rga} or even for classical statistical mixtures since it fails to take into account the correlation between the quantum and the classical degrees of freedom~\cite{UCLpost_quantum,UCLhealing}.

Nonetheless, consistent classical-quantum theories of Newtonian gravity have been studied via continuous measurement and feedback approaches \cite{Kafri_2014, Kafri_2015, tilloy2016sourcing, tilloy2017principle}, and in \cite{diosi2011gravity} using a master equation for classical-quantum dynamics \cite{blanchard1993interaction,diosi1995quantum,poulinKITP2}. These approaches are all mathematically consistent and do not suffer from the problems of the standard semi-classical approach \cite{UCLpost_quantum, UCLhealing}. Instead, the resulting master equation is linear in the combined classical-quantum state, preserves the classical-quantum split, is completely positive (CP) on the quantum system, meaning that it preserves the positivity and normalization of probabilities -- probability distributions are mapped to probability distributions. More recently, the most general form of consistent classical-quantum dynamics has been found \cite{UCLpost_quantum,UCLpawula}, which enables the construction of a fully relativistic theory of classical general relativity coupled to quantum field theory \cite{UCLpost_quantum,UCLPISHORT}. This framework has been studied in \cite{UCLunrav, UCLdec_Vs_diff, UCLconstraints, UCLhealing} and contains continuous measurement and feedback as a special case \cite{UCLhealing}. 

It was recently proven in \cite{UCLdec_Vs_diff} that in order for any Markovian classical-quantum dynamics to be completely positive - which is required for the dynamics to be consistent when acting on half an entangled state -  there must be a trade-off between the amount of decoherence on the quantum system and the amount of diffusion in the classical system. An important precursor to this result can be found in the constant force master-equation of Di\'osi \cite{diosi1995quantum}. More generally, since the trade-off can be shown to be a feature of all classical-quantum dynamics, this trade-off provides an experimental signature, not only of models of hybrid Newtonian dynamics such as \cite{diosi2011gravity}, or of post-quantum theories of General Relativity such as \cite{UCLpost_quantum}, but of \textit{any} theory which treats gravity as being fundamentally classical. The metric necessarily diffuses away from what Einstein's General Relativity predicts. This signature {\it squeezes} classical-quantum theories of gravity from both sides: if one has shorter decoherence times for superpositions of different mass distributions, one necessarily has more diffusion of the gravitational metric. One can thus use Cavendish-type experiments to upper bound the amount of diffusion, and coherence experiments to lower bound it, thus squeezing the parameter space of the theory from both sides.
The decoherence-diffusion trade-off, therefore, provides a way of testing CQ theories: one lower bounds the amount of diffusion the theory must have from coherence experiments, which can then be tested by measuring the noise in precision mass experiments. 

The present work aims to study the weak field limit of classical-quantum theories of general relativity and to provide a series of kernels that characterise correlations in the stochastic dynamics. From these, theorists and experimentalists can choose to develop and test the parameter space of CQ theories. 
In the main body, we do so in two different ways. We construct the Newtonian limit of both the path CQ general relativistic path integral of \cite{UCLPISHORT} and of the master equation from \cite{UCLpost_quantum}. The master equation is then unraveled in the form of coupled Langevin stochastic differential equations for the classical gravitational field and the quantum state of matter.
In order to do so, we first identify the relevant degree of freedom as being the Newtonian potential in a non-relativistic setting. To corroborate our results, in \cite{UCLNordstrom}, we verify our approach by comparing it to a diffeomorphism invariant CQ theory of scalar gravity, which gives the correct Newtonian limit.

If such theories are in violation of experiments, this will provide an indirect test for the quantum nature of gravity. In \cite{UCLdec_Vs_diff}, the decoherence diffusion trade-off was used to rule out a large class of natural theories that we derive here, namely those which are ultra-local, non-relativistic and continuous in the classical phase space. Hence, an important problem is to study and classify consistent classical-quantum theories of gravity, and their low energy limit, in order to further squeeze the parameter space of physically sensible CQ theories.

Moving away from fundamentally classical fields, we also mention that CQ theories of gravity could describe an effective regime of quantum gravity whenever the gravitational field freedom behaves classically. In this case, we expect that variants of the master equation and path integral we find here will be useful in describing this limit. However, an effective theory will be non-Markovian in some regimes, meaning that the decoherence-diffusion trade-off will not need to hold for all times \cite{UCLpost_quantum}. Lastly, given that we are discussing a semiclassical model of Newtonian gravity, in the discussion we compare and contrast our results with those of \cite{ruffini1969systems,diosi2011gravity,kafri2014classical,tilloy2016sourcing,tilloy2017principle}. Here, we can arrive at the master equation formulation and path-integral, from a reduction of the degrees of freedom of a fully relativistic theory. We find that our weak-field limit shares some features of \cite{diosi2011gravity,kafri2014classical,tilloy2016sourcing,tilloy2017principle}, but differs from them in several ways.

Let us now summarize our main results and present the outline of the paper.

\subsection*{Summary of results}

Section \ref{sec: CQdyn}: We start by recapping the CQ theory framework in its most general form. We explain how the state of the system is given by a hybrid CQ state $\cqstate(z)$ that can be evolved either through a path integral \eqref{eq: transition}, which was introduced in \cite{UCLPILONG, UCLPISHORT}, via a master equation \eqref{eq: CQcontinuous} as in the original formulation \cite{UCLpost_quantum} or via its unravelling \cite{UCLhealing,UCLunrav}. The evolution is completely positive and trace-preserving. We outline the degrees of freedom playing the role of the classical and quantum systems.

Section \ref{sec: Newtlim}: We review a route to the Newtonian limit of classical gravity through a reduced action approach. 
We start from a reduced action by identifying the true degree of freedom of the non-relativistic Newtonian limit as the scalar perturbation of the metric, arriving at the Newtonian ADM Hamiltonian (\ref{eq: reducedHam}). This result will allow us to construct a hybrid gravitational system in the Newtonian limit in the master equation picture, where implementing the full GR constraint is challenging~\cite{UCLconstraints}. For reference and comparison, in Appendix \ref{app: NewtLimitFromGR} we also summarise the standard derivation of the Newtonian limit which uses the full Einstein's equations. Lastly, we present a stochastic classical analogue of the CQ theory for Newtonian gravity, where the field is sourced by a Markovian noise process. We see the role of the shift vector in the imposition of stochastic Newtonian constraints. In the end, it will turn out that this is the \textit{actual classical limit} of the CQ theory when the matter degrees of freedom have completely decohered, and only the noise process remains in the classical degrees of freedom.

Section \ref{sec: PIcq}: We summarise the path integral formalism for the evolution of the CQ state $\varrho(z)$. Trajectories of the state between initial and final times are weighted by a CQ action $\mathcal{I}_{CQ}$ that defines completely positive dynamics. The natural class of theories coming from a covariant CQ proto-action introduced in \cite{UCLPISHORT} which we summarise. Treating general relativity and its Newtonian limit in the CQ framework is done through this reduced proto-action approach.

Section \ref{sec: fulltheoryNL}: We derive the CQ Newtonian limit path integral as a gauge-fixed, non-relativistic limit of the diffeomorphism invariant CQ path integral for general relativity. The gauge fixing is informed by the reduced phase space approach to the Newtonian limit, and the non-relativistic limit is implemented by keeping leading terms in  the speed of light $c$. This is one of the main results of this work. The dynamics is CP on the subset of states defined by the Newtonian limit. We find a generic prediction of CQ theories. The Newtonian potential diffuses away from its classical solution by an amount that depends on the decoherence rate into mass eigenstates.

Section \ref{sec: MasterEQ}: In this section, we introduce the master equation formalism for the evolution of the CQ state and link it to the decoherence-diffusion trade-off. 
Here, we are concerned with the \textit{continuous} form of the master equation. For an in-depth study of the continuous and jumping ME, we refer the reader to \cite{UCLpawula}. We also briefly mention how CQ dynamics can be unravelled as a system of coupled stochastic differential equations for the classical and quantum degrees of freedom \cite{UCLunrav,UCLhealing}.

Section \ref{sec: CQnewtlim}: After having introduced the master equation, here we construct the weak field regime for the class of master equations with continuous back-reaction on the gravitational degrees of freedom (\ref{eq: contbackreaction}). The effects of the dynamics are parameterized by a handful of functional parameters which can be squeezed from experiments via the decoherence diffusion trade-off. In order to discuss experimental bounds, it is necessary to impose gravitational constraints on the CQ evolution. The constrained master equation is then unravelled to derive one of the main results (\ref{eq: constraintfinal}), determining the trajectory of any CQ state through its hybrid phase space. 

We review how such theories are testable~\cite{UCLdec_Vs_diff}. The most striking being the findings of \cite{UCLdec_Vs_diff} that classical-quantum theories of gravity, which are continuous in the gravitational degrees of freedom and produce only ultra-local, non-relativistic correlations, are already ruled out. Furthermore when the coupling constants $D_0, D_2$ are constant kernels, we arrive at the Newtonian theory considered in \cite{tilloy2016sourcing, tilloy2017principle}. If one tried to minimise the amount of decoherence \cite{tilloy2017principle}, one finds that theories with constant couplings are in tension with heating experiments if the Newtonian approximation is valid below scales of $10^{-15}m$ \cite{tilloy2017principle}. This calls for a study of relativistic corrections to hybrid theories.

Section \ref{sec: discussion}: We compare and contrast with other models of semiclassical Newtonian gravity and explain the bridge with the work of \cite{diosi2011gravity,tilloy2016sourcing} while highlighting the difference with previous measurement feedback and collapse models. We summarise how our main results have been cross-checked through the use of a variety of different methods. We conclude with a discussion of our results and comments on the theoretical and experimental challenges which remain open in constructing and testing theories with a classical gravitational field coupled with quantum matter. 

\section{Classical-quantum dynamics}\label{sec: CQdyn}

In this section, we summarise the formalism used to describe the general form of consistent coupling between classical and quantum degrees of freedom~\cite{UCLpost_quantum,UCLpawula}. Early examples of these dynamics include the works of \cite{blanchard1993interaction,blanchard1995event,diosi1995quantum,alicki2003completely,poulinKITP2}. The state of the entire physical system will take the form of a hybrid classical-quantum object, while the dynamics can be expressed in three different formalisms: a master equation, the Langevin equations derived from its unravelling, and the path integral. Regardless of the choice of formalism, the dynamics are linear in the density matrix, completely positive and trace-preserving.

We take the classical degrees of freedom to live in a classical configuration space $\mathcal{M}$, and we generically denote elements of the classical space by $z$. For example, we could take the classical degrees of freedom to be position and momenta, in which case $\mathcal{M}= (\mathbb{R}^{2})$. The quantum degrees of freedom are described by a Hilbert space $\mathcal{H}$.  Given the Hilbert space, we denote the set of positive semi-definite operators as $S_{ }(\mathcal{H})$. Then, the CQ object defining the state of the CQ system at a given time is a map $\cqstate : \mathcal{M} \to  S_{}(\mathcal{H})$ subject to a normalization constraint $\int_{\mathcal{M}} dz \Tr{\mathcal{H}}{\cqstate} =1$. To put it differently, we associate to each classical degree of freedom an un-normalized density operator $\cqstate(z)$ such that $\Tr{\mathcal{H}}{\cqstate} = p(z) \geq 0$ is a normalized probability distribution over the classical degrees of freedom and $\int_{\mathcal{M}} dz \cqstate(z) $ is a normalized density operator on $\mathcal{H}$.

Classical-quantum states always admit a decomposition $\varrho(z,t) =p(z,t) \hat{\sigma}(z,t)$ where $\hat{\sigma}(z,t)$ is a normalized quantum state. Intuitively, $\hat{\sigma}(z,t)$ can be understood as the quantum state one assigns to the system, given the classical state $z$ is observed. Since the density matrix has a statistical foundation, $p(z,t)$ is then understood as the probability (density) of being in the classical state $z$. 

An example of such a {\it CQ-state} is the CQ qubit, where we take a 2-dimensional Hilbert space and couple it to classical position, and momenta \cite{UCLpost_quantum, UCLunrav}. The state then takes the form of a  $2\times 2$ matrix over phase space 
\begin{equation}
\cqstate(q, p, t)=\left(\begin{array}{ll}
u_{0}(q, p, t) & c(q, p, t) \\
c^{\star}(q, p, t) & u_{1}(q, p, t)
\end{array}\right).
\label{eq:cqqubitBackground}
\end{equation}

If one desires to treat gravity coupled to matter fields in the CQ framework, a natural choice for the classical degree of freedom is the metric, while the matter fields are quantised and play the role of quantum degrees of freedom. This allows for proper treatment of semiclassical gravity, where correlations between the gravitational field and matter are not ignored as in the semiclassical Einstein equations. However, from the path integral formulation, we can see that the positivity of the full Einstein equations in the CQ framework may not be ensured for every possible initial state \cite{UCLPISHORT}. Nevertheless, as discussed in Section \ref{sec: fulltheoryNL}, when we restrict to the Newtonian limit, 
positivity is ensured. This implies that the final picture sees the role of the classical degree of freedom being played by the scalar perturbation $\Phi$, while the matter degrees of freedom are quantum. In this paper, the matter is chosen to take the form of pressureless dust, and the associated mass density operator will be indicated by $\hat{m}(x)$, which should not be confused with the mass itself.

One desires the dynamics of these states to retain their positivity, preserve the statistical interpretation of the density matrix and give rise to positive probabilities when acting on half-entangled states. Therefore, the dynamics must be linear, completely positive (CP) and probability-preserving. If the dynamics are also time-local, then consistent CQ dynamics can be written in the form of a path integral \cite{UCLPILONG,UCLPISHORT}, master equation \cite{UCLpost_quantum} or its unravelling in terms of stochastic differential equations \cite{UCLhealing,UCLunrav}.

In this paper, we will present the Newtonian limit of CQ gravity in all these forms. After the next section, readers more used to dealing with path integrals should keep reading as we will introduce the CQ path integral formalism in Section \ref{sec: PIcq} and derive the Newtonian limits right after by gauge fixing the full diffeomorphism invariant CQ theory of general relativity. 
If instead the reader feels more comfortable with master equations and unravellings we suggest that, after the next section, you could skip to Section \ref{sec: MasterEQ}. There, we review the CQ master equation formalism and explain how it can be unravelled to examine single trajectories of the CQ state. This will help to better appreciate one of the central results (\ref{eq: constraintfinal}) of the paper, which is presented in the unravelling formalism.

In the next section, we will derive the Newtonian limit of classical GR, starting from the reduction of the degrees of freedom to scalar perturbations, in the Hamiltonian formalism, which will then be key when discussing the Newtonian limit of CQ gravity.

\section{Newtonian limit of classical GR}\label{sec: Newtlim}

In this section, we study the weak field and Newtonian limits of classical general relativity (GR), which motivates our study of the weak field and Newtonian limits of classical-quantum theories of gravity. By the weak field limit, we mean the linearised expansion of the metric around a flat Minkowski background, while by the Newtonian limit, we refer to a non-relativistic setting by taking the $c\rightarrow\infty$ limit and discarding terms with high powers of inverse $c$.

The Newtonian limit of GR is represented by a non-dynamical scalar perturbation of flat Minkowski spacetime expressed through the metric:
\begin{equation}\label{eq: Newtonianmetricapprox}
        ds^2=-c^2\left(1+\frac{2\Phi}{c^2}\right)dt^2+\left(1-\frac{2 \Phi}{c^2}\right)\delta_{ij} dx^i dx^j ,
    \end{equation}
where $\Phi$ satisfies the gravitational Poisson equation. 
As a reminder, we present the usual derivation of this limit from a gauge fixing of the full Einstein theory in Appendix \ref{app: NewtLimitFromGR}. There, we start from a generalized scalar-vector tensor perturbation of the metric in the form of
\begin{equation}
        ds^2=-c^2\left(1+\frac{2\Phi}{c^2}\right)dt^2+ \frac{w_i}{c}(dt dx^i+dx^i dt)+\left[\left(1-\frac{2\psi}{c^2}\right)\delta_{ij}+\frac{2s_{ij}}{c^2}\right] dx^i dx^j,
    \end{equation}
where $\partial_iw^i=\partial_is^{ij}=0$ and we take the infinite $c$ limit of Einstein's equations. When the stress-energy tensor is chosen to represent a pressureless dust distribution, only one non-dynamical scalar perturbation $\Phi$ remains at the end, and it is constrained to obey the gravitational Poisson's equation.

Based on the knowledge obtained from the full GR derivation, we instead present a derivation of the Hamiltonian formulation of the Newtonian limit, which starts directly from a reduction of the degrees of freedom to scalar perturbations.  In the reduced degrees of freedom approach, we \textit{first} assume that the relevant degrees of freedom are scalar perturbations. We shall also allow for vector perturbations at higher order in $c$, which we find are necessary to construct a consistent CQ theory. 

As we will show in the rest of the main body, this provides us with a way of constructing the Newtonian $c\to \infty$, $\pi_{\Phi} \approx 0$ limit of CQ theories via a reduction of the gravitational degrees of freedom, even in the absence of a complete CQ theory of GR which is positive on all possible states. This is explained in detail in Section \ref{sec: fulltheoryNL}, where we show that the problematic terms appearing in the CQ treatment of GR vanish in the Newtonian limit, validating the limit with a top-down approach. Regardless, we get the same results in both the path integral and the master equation approaches.

\subsection{Newtonian limit via a scalar reduced action}\label{ss: NewtLimitFromAction}
To derive the Newtonian limit of GR via a reduced Hamiltonian, we take as a starting point the linearised Einstein Hilbert Lagrangian density, which is equivalent to the spin-2 field Fierz-Pauli action \cite{Fierz} for the metric perturbation $g_{\mu \nu} = \eta_{\mu \nu} + h_{\mu \nu}$:
    \begin{align} \label{eq: linaction}
        &\mathcal{S}[h_{\mu\nu}]=\frac{c^4}{16\pi G}\int d^4x \, \mathcal{L}(h_{\mu\nu}), \\
        &\mathcal{L}(h_{\mu\nu})= -\frac{1}{2}\partial_\mu h^{\mu\nu}\partial_\nu h+ \frac{1}{2}\partial_\mu h^{\rho\sigma}\partial_\rho h^\mu_\sigma-\frac{1}{4}\eta^{\mu\nu}\partial_\mu h^{\rho\sigma}\partial_\nu h_{\rho\sigma}+\frac{1}{4}\eta^{\mu\nu}\partial_\mu h\partial_\nu h.
    \end{align}
We are interested in constructing CQ dynamics for a Newtonian theory, so we further make a Newtonian approximation of the metric. We take the ADM decomposition 
\begin{equation}
\mathrm{d} s^{2}=-(N c \mathrm{~d} t)^{2}+g_{i j}\left(\mathrm{~d} x^{i}+N^{i} c \mathrm{~d} t\right)\left(\mathrm{d} x^{j}+N^{j} c \mathrm{~d} t\right),
\end{equation}
and make the weak field assumption 
\begin{equation} \label{eq: approxMetric}
    N=\left(1+\frac{\Phi}{c^2}\right),\  N^i =\left( 0 + \frac{n^i}{c^3}\right), \ g_{ij} =\left ( 1-\frac{2 \psi}{c^2}\right)\delta_{ij}.
\end{equation}
The extra factor of $c$ in the choice of shift-vector is related to the fact that, classically, the $h_{0i}$ component occurs at a higher order than the $h_{00}$, $h_{ij}$ components \cite{CarrollNotes}. We assume that all fields vanish at infinity. In the purely classical case, we find that when the stress-energy tensor $T_{0i}=0$, then $n^i=0$, but we will show that in the combined CQ case $n^i \neq 0$ even in the absence of the stress-energy tensor. Instead, a non-zero $n^i$ is required in order to preserve the Hamiltonian constraint of the theory. 

With the gauge fixing of Equation \eqref{eq: approxMetric}, the linearized action in Equation \eqref{eq: linaction} is 
\begin{equation}
\label{eq: newtaction}
\begin{split}
    \mathcal{S}=\frac{1}{8 \pi G }\int d^4 x &\bigg[-\frac{3 \dot{\psi}^2}{c^2} + \frac{\partial_i n^i}{2c^2}  ( \dot{\Phi}- \dot{\psi}) - \frac{\dot{n}^i}{2c^2}  (\partial_i \Phi + 3\partial_i \psi ) -\frac{1}{4c^2} \partial_i n^j \partial_j n^i \\&\quad + \partial_i \psi \partial^i \psi - 2 \partial_i \Phi \partial^i \psi + \frac{1}{4c^2} \partial_i n^j \partial_j n^i \bigg].
\end{split}
\end{equation}

To go to the Hamiltonian picture, we first calculate the functional derivatives with respect to $\dot{\psi}$, $\dot{\phi}$ and $\dot{n}^i$ to find the conjugate momenta
  \begin{equation}\label{eq: conjmomenta}
    \pi_{\psi} = -\frac{1}{16 c^2 G \pi }(12\dot{\psi}+\partial_i n^i), \ \pi_{\Phi} = \frac{\partial_i n^i }{16 \pi c^2 G}, \  \pi_{i} = - \frac{1}{16 \pi c^2 G}(\partial_i\Phi+3 \partial_i \psi). 
    \end{equation}  
We see that Equation \eqref{eq: conjmomenta} defines two primary constraints. As a reminder, in Hamiltonian mechanics, a primary constraint is a relation between a coordinate and its conjugate momenta that holds regardless of the equation of motion. Here, the equations for $\pi_{\Phi}$ and $\pi_{i}$ are constraints because $\dot{\pi}_{\Phi}$ and $\dot{\pi}_{i}$ will depend only on the initial state i.e. first order derivatives of the canonical coordinates, while only the evolution equation for $\pi_{\psi}$ will depend on second order derivatives. We thus have two primary constraints:
    \begin{align} \label{eq: primaryConstraint}
    &\Pi_\Phi=\pi_\Phi - \frac{\partial_i n^i }{16 \pi c^2 G} \approx 0, \\
    &\label{eq: primaryConstraint2} \Pi_i = \pi_i + \frac{1}{16 \pi c^2 G}(\partial_i\Phi+3 \partial_i \psi) \approx 0,
    \end{align}
where $\approx 0$ means \textit{weakly zero} in the Dirac sense \cite{diracbook}, that the quantity is set to $0$ by constraining the initial data. In the $c \to \infty$ limit, Equations (\ref{eq: primaryConstraint}\,-\ref{eq: primaryConstraint2}) become the constraints $\pi_{\Phi}, \pi_i \approx 0$, which enforce the usual Hamiltonian and momentum constraints.

One might worry about the fact that the constraints (\ref{eq: primaryConstraint}\,-\ref{eq: primaryConstraint2}) derived from the (Fierz-Pauli) linearised gravity action appear to be different with respect to the constraints
one obtains by first considering the constraints in the full ADM formalism and then linearising them.
This has been studied in \cite{Green:2008qa}, where it was shown that the two forms are simply related by a canonical transformation.
Alternatively, one could follow the approach of \cite{diracbook} and add a specific non-covariant term to the linearised action \eqref{eq: linaction}.
This term vanishes on shell and simplifies the primary constraints to match them with those derived from the ADM formalism.

Since we are interested in the $c\to \infty$ limit, these distinctions do not matter, as we end up imposing the constraints $\pi_{\Phi}, \pi_i \approx 0$, which are equivalent to the primary constraints $\pi_{N}, \pi_{N_i} \approx 0 $, where $N$, $N^i$ are the lapse and shift vectors.

Using the definitions of conjugate momenta in Equation \eqref{eq: conjmomenta} and working to leading order in $c$, we arrive at the Newtonian Hamiltonian 

\begin{equation}
\label{eq: reducedHam}
    H^{(gr)}=\int d^3x \left[-\frac{2\pi Gc^2}{3}\pi_\psi^2  -\frac{1}{12} \pi_{\psi} \partial_i n^i  +\frac{\partial_i \psi \partial_i \Phi}{4\pi G}-\frac{\partial_i \psi \partial_i \psi }{8\pi G} + \lambda_{\Phi}\pi_{\Phi} + \lambda^i \pi_i \right].
\end{equation}

To find the Newtonian interaction Hamiltonian, we need to couple gravity to matter. When a matter action $S_m$ is included, the coupling to the linear perturbation is found via $h_{\mu \nu} T^{\mu \nu}$, which is required to reproduce Einstein's equations. We shall consider the matter distribution to be that of a particle with mass density $m(x)$ and, because we are working in the non-relativistic limit, we shall assume that only $T_{00}$ acts as a source for the gravitational field. The corresponding interaction Hamiltonian can be then easily written as
    \begin{equation}\label{eq: intHam}
        H^{I}=\int d^3x\,\mathcal{H}^{I}=\int d^3x\, \Phi(x)m(x),
    \end{equation}
such that the total Hamiltonian is given by $H_{tot}=H^{(gr)}+H^{I}$:
    \begin{equation}
        \label{eq: totham}
        H_{tot}=\int d^3x \left[-\frac{2\pi Gc^2}{3}\pi_\psi^2  -\frac{1}{12} \pi_{\psi} \partial_i n^i  +\frac{\partial_i \psi \partial_i \Phi}{4\pi G}-\frac{\partial_i \psi \partial_i \psi }{8\pi G} + \lambda_{\Phi}\pi_{\Phi} + \lambda^i \pi_i + \Phi(x)m(x)\right].
    \end{equation}
The dynamics associated with $H_{tot}$ can be derived from Hamilton's equations and is given by:
\begin{equation}\label{eq: newtlimitdyn}
     \dot{\psi} = - \frac{4G \pi c^2 \pi_{\psi}}{3} - \frac{1}{12} \partial_i n^i, \ \dot{\pi}_{\psi} = \frac{\nabla^2( \Phi - \psi)}{4\pi G}, \ \dot{\Phi} = \lambda_{\Phi}, \  \dot{\pi}_{\Phi}= \frac{\nabla^2 \Phi}{4 \pi G } -m , \ \dot{n}^i = \lambda^i, \  \dot{\pi}_i = -\frac{1}{12} \partial_i \pi_{\psi}
\end{equation}
and we arrive at Newtonian gravity by solving Equation \eqref{eq: newtlimitdyn} since the constraint $\pi_{\Phi} \approx 0$ imposes 
    \begin{equation}
        \frac{\nabla^2 \Phi}{4 \pi G } -m \approx 0 \quad \Rightarrow \quad \Phi(t,x) =-G \int d^3 x'\; \frac{ m(x')}{|x-x'|} 
    \end{equation}
on the potential $\Phi$ i.e, $\Phi$ must solve Poisson's equation. On the other hand, the constraint $\pi_i \approx 0 $ imposes  $\pi_{\psi} \approx 0$, where we have used the fact that $\pi_{\psi}$ vanishes at infinity. Preservation of the $\pi_{\psi} \approx 0$ constraint imposes that $\Phi = \psi$. Moreover, the time derivative of the Newtonian potential directly dictates the Lagrange multiplier via $\lambda_{\Phi}=\dot{\Phi}$ and the divergence part of the shift vector via:
    \begin{equation}
        \dot{\Phi} = - \frac{1}{12} \partial_i n^i .
    \label{eq:philapse}
    \end{equation}
Since we assume a stationary source, where only $T_{00}$ contributes, this imposes that $\partial_i n^i =0$. Note this does not entirely fix the shift $n^i$, and solutions related by different choices of the shift vector will be gauge equivalent. In the classical theory, it is common to assume the gauge $n^i=0$, in which case we arrive at the Newtonian metric of Equation \eqref{eq: Newtonianmetricapprox}, where $\Phi$ satisfies Poisson's equation.

We have arrived at the Newtonian limit of general relativity by making the Newtonian approximation on the metric in Equation \eqref{eq: approxMetric} and then deriving the dynamics in the $c\to\infty$ limit. While deriving the Newtonian limit from a full GR approach requires a complete diffeomorphism invariant theory, we have seen that we can construct a consistent reduced theory by first identifying the correct degrees of freedom (in this case, scalar perturbations of the metric) and then writing down their dynamics according to a reduced Hamiltonian.

Before discussing how a quantum system's back-reaction on the classical Newtonian field is implemented through diffusion processes, we would like to mention our choice of gauge. The end goal of this work is to formulate the Newtonian limit of gravity for CQ-hybrid theories; we do not know if a complete CQ theory can be made fully diffeomorphism invariant. Regardless, our choice of gauge is motivated by the need to preserve the gravitational constraints. By choosing the gauge (i.e. coordinates) as in (\ref{eq: approxMetric}), we know that we have a way of consistently selecting trajectories that stay on the constraint surface, where the conjugate momenta vanish as described in this section.

\subsection{The weak field classical limit}

We will first consider a purely stochastic modification to classical general relativity. With the benefit of hindsight, it corresponds to taking the classical limit of the matter fields in the CQ theory. However we present it first, partly for simplicity, and partly because for some experiments, it is the regime of interest. It also provides an interesting analogy with quantum gravity. Note that although this limit is a stochastic theory of classical gravity, it is different to what is usually referred to as {\it stochastic gravity} \cite{hu2008stochastic} which is based off of the semi-classical Einstein's equations. For example, the theory here is Markovian, while stochastic gravity needs to be non-Markovian in the case of statistical mixtures of states with a large variance.

In the CQ case, we will construct the Newtonian limit by taking the non-relativistic limit, which means that the relevant degrees of freedom are scalar perturbations of the metric of the form in Equation \eqref{eq: approxMetric} and then considering a reduced CQ master equation governing the dynamics of the perturbations. Since we will be interested in describing the non-relativistic limit of a quantum mass interacting with classical gravity, the back-reaction on the gravitational field from the quantum matter is dominated by the $T_{00}$ component, and any classical-quantum momentum constraint \cite{UCLconstraints} will be unchanged since it does not involve matter. In particular, the back-reaction of the quantum system on the classical system enters through the  $\pi_{\Phi}$ equation in \eqref{eq: newtlimitdyn}. Loosely speaking, because quantum back-reaction must necessarily involve diffusion, the equation of motion for $\dot{\pi}_{\Phi}$ will be modified to include a stochastic term. To gain some intuition, we can consider the classical analogue by considering a Langevin equation for $\dot{\pi}_{\Phi}$
\begin{equation}\label{eq: hamconst}
   \dot{\pi}_{\Phi}= \frac{\nabla^2 \Phi}{4 \pi G } -m - \sigma\xi ,
\end{equation}
where $\sigma(x)$ is a coefficient and $\xi(t,x)$ is a white noise process which we will relate to the $D_2$ coefficient appearing in the CQ master equation \eqref{eq: cqdyngen}. We will later find that this stochasticity is all that is required to maintain complete positivity in the CQ case. With the modified dynamics for $\pi_{\Phi}$, we find the constraint $\pi_{\Phi} \approx 0$ imposes $\frac{\nabla^2 \Phi}{4 \pi G } = m +  \sigma \xi $ on the potential $\Phi$. The momentum constraint $\pi_i$ remains unchanged from the deterministic case, and its preservation again imposes the constraint $\Phi = \psi$. However, with the addition of the noise, the solution to the Newtonian constraint is no longer stationary but is instead given by a solution to the random Poisson equation 
    \begin{equation} 
\label{eq: randomPoissonClass}
\Phi  = -G \int dx'\; \frac{ m(x') + \sigma(x')\xi(x',t)}{|x-x'|} .
    \end{equation}
Preservation of the Hamiltonian constraint in Equation \eqref{eq: hamconst} then determines $\lambda_{\Phi} $ and $\partial_i n^i$. In particular, with the gauge choice of Equation \eqref{eq: approxMetric}, we see that $n^i$ is required in order for the theory to be consistent. The presence of diffusion, combined with the quantum back-reaction, will make the Newtonian potential $\psi = \Phi$ fluctuate, and its evolution will be determined by the shift vector $\partial_i n^i$ via Equation \eqref{eq: newtlimitdyn} or more immediately, via Equation \eqref{eq:philapse} in particular. This is one of the key 
results of this paper, and we will return to it when we discuss the master equation approach. Without allowing the shift to be a freely chosen gauge parameter, the momentum  conjugate to the Newtonian potential $\Phi$, looks like it will diffuse off to infinity via the random walk process given by Equation \eqref{eq: hamconst}.

We point out once again that this does not fix the shift $n^i$ uniquely since we are free to add a divergenceless term and get the same solution to the equation of motion. Moreover, in a complete calculation, we expect that contributions from $T_{0i}$ will also determine the components $n^i$ without affecting the Newtonian contribution, which is given by the $h_{00}$ component. Regardless, performing higher-order calculations is beyond the scope of the current work. 

In the stochastic case, we still find that $\Phi = \psi$ is set by the dynamics since the addition of noise in Equation \eqref{eq: hamconst} is the only modification to the theory. Hence, one can instead start with the metric perturbation:
\begin{equation} 
\label{eq: approxMetricSimple}
    N= \left(1+\frac{\Phi}{c^2}\right),\  N^i = \left(0 + \frac{n^i}{c^3}\right), \ g_{ij} = \left( 1-\frac{2 \Phi}{c^2}\right)\delta_{ij}.
\end{equation}
and consider the dynamics obtained by setting $\Phi=\psi$ in Equation \eqref{eq: reducedHam}. One could even remove the kinetic term $-\frac{2\pi Gc^2}{3}\pi_\psi^2$, which doesn't contribute to the equations of motion on the constraint surface, and the Lagrange multiplier term $\pi_{\Phi}$, which is redundant as the Hamiltonian is also linear in  $\pi_{\psi}$. In this case, the reduced Hamiltonian reads.
\begin{equation}\label{eq: reducedHamEquals}
    H^{(gr)}+ H^{I} 
        =\int d^3x \left[ \frac{(\nabla \Phi)^2}{8\pi G}+  m\Phi  -\frac{1}{12} \pi_{\Phi} \partial_i n^i +  \lambda^i \pi_i \right].
\end{equation}
Equations \eqref{eq: approxMetricSimple} and \eqref{eq: reducedHamEquals} are considerably simpler than Equations \eqref{eq: approxMetric} and \eqref{eq: reducedHam} but result in the same dynamics, we will use the former to describe the Newtonian limit of CQ theories. 

Note that since the noise process is a white noise process, technically, the metric perturbation will describe a probability measure, and we should use it to compute averaged quantities. This is true of both $\Phi$ and the shift vector $n^i$ which are now both stochastic quantities. In particular, averaging over a timescale $\Delta T$ and length scale $\Delta L$, $\Delta L/\Delta T \ll c$, we have that  $\frac{\Delta \Phi}{\Delta T } \sim \frac{n^i }{\Delta L }$, so that $\frac{ \Delta L \Delta \Phi}{\Delta T } \sim n^i \ll c^3 $ which verifies our initial assumption to include the perturbation $h_{0i}$ as $\frac{n^i}{c^3}$ in Equation \eqref{eq: approxMetric}.

With this in mind, we now study the Newtonian limit of the full CQ theory. We find in the $c\to \infty$ limit that we arrive at Poisson's equation on average, but because of the CQ interaction, the Newtonian limit also predicts diffusion around this solution according to Equation \eqref{eq: randomPoissonClass}, with simultaneous decoherence on the quantum system. In \cite{UCLNordstrom}, we study a diffeomorphism invariant theory of CQ scalar gravity and show that in the $c \to \infty $ limit, the results are quantitatively and qualitatively the same as in the reduced degrees of freedom approach. This gives us more confidence that our results are independent of the coordinate choice we have made.

One can see the analogy between the quantum and the stochastic cases, where the order of operations of introducing the constraint and quantizing/inserting noise matters. Unlike the quantum case, it is easier to see why the insertion of stochasticity has different effects depending on whether it is done before or after the phase space reduction. Imagine a system constrained to have spherical symmetry. If a non-spherically symmetric noise process is added, and then we project onto a spherically symmetric initial state, the noise process will continue and drive the system away from spherical symmetry unless the noise is also chosen to be spherically symmetric.  On the other hand, if the system is constrained first so that all degrees of freedom can only depend on the radius, then the noise inserted can only be spherically symmetric. We comment more about this in Section \ref{sec: discussion}.

In the next section, we introduce the path integral formalism for CQ dynamics which we will later use to derive the Newtonian limit of the full GR theory. The reader who is instead more comfortable with a master equation approach could now skip to Section~\ref{sec: MasterEQ}, where we review the master equation and unravelling formalism for the CQ dynamics and derive the Newtonian limit in that context.


\section{Path integral for CQ dynamics}
\label{sec: PIcq}

In this section, we will summarise the path integral formulation of CQ dynamics \cite{UCLPILONG, UCLPISHORT}.  While the initial theory for classical general relativity coupled to quantum matter was written in the master equation formalism \cite{UCLpost_quantum}, it is hard to frame the dynamics in an explicitly covariant manner. 
The path integral formulation, on the other hand, can make this explicit.  It can be derived from the master equation \cite{UCLPISHORT}, but it can also be taken as the starting point. As with the master equation approach, the dynamics of the hybrid system is linear in the density matrix, completely positive, and trace-preserving. 

For classical-quantum dynamics, the path integral should tell us how to evolve the components of the CQ state 
\begin{equation}\label{eq: statecomponents}
    \varrho(q,p,t) = \int d \phi^+ d\phi^- \varrho(q,p,\phi^+,\phi^-,t) \ \ket{\phi^+}\bra{\phi^-},
\end{equation}
where $q$ and $p$ stand to represent all the classical degrees of freedom and their conjugate momenta in the classical phase space, and $\phi$ are the continuous quantum degrees of freedom. Here, we bra and ket of the density matrix can be thought of as two independent fields $\phi^+$, and $\phi^-$.

Writing this out explicitly, a classical-quantum path integral will take the form \cite{UCLPISHORT, UCLPILONG}:
\begin{equation}\label{eq: transition}
    \varrho(q_f,p_f,\phi^+_f,\phi^-_f,t_f)  = \mathcal{N}\int \mathcal{D}q \mathcal{D}p\mathcal{D} \phi^+\mathcal{D}\phi^- e^{\mathcal{I}_{CQ}[q,p,\phi^+,\phi^-,t_i,t_f]}   \varrho(q_i,p_i,\phi^+_i,\phi^-_i,t_i).
\end{equation}
Where $\mathcal{I}_{CQ}$ is the CQ action and it is implicitly understood that boundary conditions are to be imposed at $t_f$. We could also include a $\mathcal{N}$ factor in case the action does not preserve the norm of the state. The path integral gives each element of the density matrix at time $t$, given an  initial density matrix.

From \cite{UCLPISHORT, UCLPILONG}, we recall that any classical-quantum configuration space path integral with action of the form \cite{UCLPISHORT}
 \begin{equation}\label{eq: positiveCQ}
 \begin{split}
  \mathcal{I}_{CQ}(q,p,\phi^+,\phi^-,t_i,t_f) &=\mathcal{I}_{CQ}^{+}(q,p,\phi^+,t_i,t_f)\\
  & + \mathcal{I}^{*-}_{CQ}(q,p,\phi^-,t_i,t_f) - \mathcal{I}_{C}(q,p, t_i,t_f),
  \end{split}
\end{equation}
defines completely positive CQ dynamics. In Equation \eqref{eq: positiveCQ} $\mathcal{I}_{CQ}^\pm$ determine the CQ interaction for the bra or ket ($\phi^\pm$) of the density matrix, and $\mathcal{I}_C(q,p, t_i,t_f)$ is a purely classical Fokker-Planck like action \cite{Kleinert, Sieberer_2016} which should be positive definite, at least for large values of the classical variables $q,p$ so that the path integral converges. A simple example of such a Fokker-Planck term describes the standard Brownian motion in phase space with classical degrees of freedom and its conjugate momentum $(q,p)$: 
 \begin{align}
   \mathcal{I}_{C}(\q,p, t_i,t_f)= -\frac{1}{2D_2}\int_{t_i}^{t_f} dt \left(\dot{p}+\frac{\partial V(q)}{\partial q}\right)^2
   +\log{\delta\left(\dot{q}+\frac{p}{m}\right)}
   \label{eq:fpaction}
 \end{align}
which will allow for diffusion away from the deterministic force law given in terms of the purely classical potential term $V(q)$ coming from Hamilton's equations, while suppressing paths that are further away from the average force, by an amount depending on the diffusion coefficient $D_2$. In this example, we are working in phase space rather than configuration space, with $\delta$-function keeping the relationship between the momentum $p$ and $\dot{q}$.

On the other hand, a simple example of  $\mathcal{I}_{CQ}$ is 
  \begin{align}
   \mathcal{I}_{CQ}(q,p,\phi^+, t_i,t_f)+\mathcal{I}_{CQ}(q,p,\phi^-, t_i,t_f)=- \frac{1}{2D_2}\int_{t_i}^{t_f} dt &\bigg[ \left(
   \dot{p}
   +\frac{1}{2}\frac{\partial V_{CQ}(q,\phi^+)}{\partial q}
   +\frac{1}{2}\frac{\partial V_{CQ}(q,\phi^-)
   }{\partial q}\right)^2
   \nonumber\\
   &+\frac{1}{8D_2}\left(\frac{\partial V(q,\phi^-)
   }{\partial q}-\frac{\partial V(q,\phi^+)}{\partial q}\right)^2\bigg]
   \label{eq:cqactionexample}
 \end{align}
where the first line of the CQ action looks like the Fokker-Planck action of Equation \eqref{eq:fpaction} but sourced by the average of a CQ potential, with the average taken of the bra and ket quantum degree of freedom. I.e. by 
 \begin{align}
 \bar{W}_{CQ}:=\frac{1}{2}\left(\frac{\partial V_{CQ}(q,\phi^+)}{\partial q}
   +\frac{\partial V_{CQ}(q,\phi^-)}{\partial q}\right)\,.
   \label{eq:protoexample}
 \end{align}  
The second term in Equation \eqref{eq:cqactionexample} is a Feynman-Vernon \cite{FeynmanVernon} term which causes decoherence by suppressing off-diagonal elements of the density matrix. One can verify that the coupling between the bra and the ket fields cancel in this case.
 
One can also go directly from the master equation picture in Equation \eqref{eq: CQcontinuous} to a path integral picture whenever the master equation contains terms that are no more than quadratic in quantum momentum operators and classical derivative operators \cite{UCLPILONG}. In this sense, Equation \eqref{eq: positiveCQ} allows for more general path integrals since one can include couplings that are higher than quadratic. In this case, the mapping between master Equations and path integrals will not always be clear. The reduced phase space weak limit of gravitational CQ that we consider in Section \ref{sec: fulltheoryNL} is of this type since, as we will see, imposing the constraint generates coupling that higher than second order.
 
Due to the fact that Equation \eqref{eq: positiveCQ} contains no $\phi^{\pm}$ cross terms, the path integrals preserve purity on the quantum system. Pure quantum states are mapped to pure quantum states, and no information is lost. Moreover, conditioned on the classical trajectory, the quantum state evolution is deterministic, which provides a natural mechanism for wave-function collapse if the classical degree of freedom is taken to be fundamental. Furthermore, because the classical degree of freedom is itself dynamical, unlike in spontaneous collapse models \cite{grw85, Pearle:1988uh, PhysRevA.42.78, BassiCollapse, PhysRevD.34.470}, it is possible to make the dynamics covariant \cite{UCLPISHORT}. We shall see an explicit example of diffeomorphism invariant CQ dynamics in Section \ref{sec: fulltheoryNL}.

The path integral constructs a CQ state at time $t_f$ from a CQ state at time $t_i$. However, we are often interested in computing correlation functions for stationary states, and we would like to obtain information on correlation functions over arbitrary long times by taking the limit $t_i \to -\infty, t_f \to \infty$. In open systems, as well as when calculating scattering amplitudes, it is often assumed that the initial state in the infinite past does not affect the stationary state of the system so that there is a complete loss of memory of the initial state \cite{Sieberer_2016}. Under this assumption, it is possible to ignore the  boundary term containing the initial CQ state $\cqstate(q_i,\phi^\pm_i, t_i)$, arriving at the partition function  
\begin{equation}\label{eq: CQpartitionfunction}
   \mathcal{Z}[J^+,J^-,J_q] = \mathcal{N}  \int\mathcal{D}q  \mathcal{D} \phi^\pm   \ e^{ \mathcal{I}_{CQ}( q,\phi^+,\phi^-,-\infty,\infty) -i( J_+ \phi^+ -J_- \phi^-) -J_q q}
\end{equation}
such that we can then use standard perturbation methods for computing correlation functions in CQ theories. For simplicity, we shall consider just the path integral partition function without the presence of sources, but all of our equations arise from studying the CQ action. 

A natural class of theories introduced in \cite{UCLPISHORT, UCLPILONG}, which work especially well for classical and quantum fields are those derivable from a classical-quantum proto-action. This we define as
    \begin{equation}
        W_{CQ}[q,\phi]=\int dtd\vec{x}\,\big(\mathcal{L}_c[q(x)]-\mathcal{V}_I[q(x),\phi(x)]\big)
    \end{equation}
where $\mathcal{L}_c$ is the Lagrangian density of the classical action $\mathcal{S}_c[q]=\int dtd\vec{x}\,\mathcal{L}_c[q]$, $\mathcal{V}_I$ is the interaction potential density $V_I[q,\phi]=\int dtd\vec{x}\,\mathcal{V}_I[q,\phi]$ and the classical and quantum degrees of freedom are now locally dependent on spacetime: 
\begin{equation}
\label{eq: fieldConfigurationSpace2}
    \begin{split}
      \mathcal{I}(q,\phi^+,\phi^-,t_i,t_f) =\int_{t_i}^{t_f} dt d\vec{x}\bigg[& i\mathcal{L}_Q[q,\phi^+]-i\mathcal{L}_Q[q,\phi^-] -\frac{1}{2}\frac{ \delta  \Delta W_{CQ}}{ \delta q_{i}} D_{0,ij}[q(x)] \frac{\delta \Delta W_{CQ}}{ \delta q_{j}} \\&- \frac{1}{2} \frac{\delta \bar{W}_{CQ} }{ \delta q_i} D_{2, ij}^{-1}[q(x)] \frac{\delta \bar{W}_{CQ} }{ \delta q_j}
    \bigg].
    \end{split}
\end{equation}
The example of Equation \eqref{eq:cqactionexample} is a special case of this, while here we define a more general form via a {\it proto action} \cite{UCLPILONG}
\begin{align}
    \Delta W_{CQ}[q,\phi^+,\phi^-]\coloneqq W_{CQ}[q,\phi^+]-W_{CQ}[q,\phi^-],
\end{align}
and a more general {\it average force} in comparison with Equation \eqref{eq:protoexample}, given by
\begin{align}
    \bar{W}_{CQ}[q,\phi^+,\phi^-]\coloneqq \frac{1}{2}\big(W_{CQ}[q,\phi^+]+W_{CQ}[q,\phi^-]\big).
\end{align}

To better understand the effect of the CQ action, we can again read Equation~\eqref{eq: fieldConfigurationSpace2} in order. The first two terms represent the unitary evolution of the quantum degrees of freedom, which can be any quantum field theory Lagrangian. Given that we are evolving a density matrix we have that the left/right (bra/ket) branches are evolved separately by the $\pm$ terms. One can also add friction terms to Equation \eqref{eq: fieldConfigurationSpace2} though we shall not do this in the present work.
Next, we find the decoherence term. It has a decoherence coefficient $D_0$ and is constructed from the variation of the difference between the left and right branches of the proto-action $\Delta W_{CQ}[q,\phi^+,\phi^-]$.
This term is responsible for the decoherence of the system, penalising trajectories of the hybrid state that move further away from $\phi^+=\phi^-$. This term does not affect the diagonal terms in the density matrix but does
suppress the off-diagonal terms exponentially with time.
Lastly, we have the diffusion term. Built from the variation of the $\pm$ averaged interaction $\bar{W}_{CQ}[q,\phi^+,\phi^-]$, this term has diffusion coefficient $D_2$ and penalises trajectories in which the classical degrees of freedom tend to deviate from their Euler-Lagrange equations of motion. In other words, the noise introduced in the classical degrees of freedom from the backreaction of the quantum degrees of freedom introduces a diffusion process that is visible from this term.

The coefficients $D_0, D_2$ need to be positive, and in the case of several classical degrees of freedom or fields positive definite matrices and kernels. To ensure that the action takes the form of Equation \eqref{eq: positiveCQ} and ensure complete positivity of the dynamics, we can impose the matrix inequality: 
    \begin{equation}
    \label{eq: decdiffsat}
        4D_2 \succeq D_0^{-1}.
    \end{equation}
This we call the \textit{decoherence-diffusion trade-off}. The trade-off itself originates from positivity conditions on the master equation, on which we expand in Section~\ref{sec: MasterEQ}, but was derived and explored in depth in \cite{UCLdec_Vs_diff}. 
The physical meaning of this relationship between the decoherence and the diffusion coefficients tells us that, if we want to preserve the coherence of the quantum degrees of freedom for prolonged times, we will have a lot of noise introduced in our classical degrees of freedom. When the trade-off is saturated, meaning that $4D_2=D_0^{-1}$, the path integral also preserves the purity of the quantum system. This form of the action is motivated by the study of path integrals \cite{UCLPILONG} for CQ master Equations, which are approximately Hamiltonian \cite{diosi2011gravity, UCLpost_quantum, UCLhealing}.  

In the next section, we gauge fix the CQ path integral for general relativity coupled with quantum matter and then take the non-relativistic limit to derive the Newtonian limit for the CQ theory.



\section{Newtonian limit from the gauge fixing of classical-quantum general relativity}
\label{sec: fulltheoryNL}

In this section, we derive the Newtonian limit of the full general relativistic diffeomorphism invariant CQ path integral which was first introduced in \cite{UCLPISHORT}, here reported as Equation~\eqref{eq:PQG-actionNewton}. To derive the Newtonian limit, we will start by gauge fixing the classical degrees of freedom using the gauge described in Equation~\eqref{eq: approxMetricSimple}, which is justified through the arguments of Section~\ref{sec: Newtlim}, and then by taking the non-relativistic limit through keeping only leading terms in the speed of light $c$.

This derivation leads to the same unravelled Newtonian limit that we will later obtain from the master equation in Equation~\eqref{eq: constraintfinal} and acts as a sanity check for the theory introduced \cite{UCLPISHORT}, showing that its constraints have a sensible non-relativistic limit. In the Newtonian limit, once the choice is made of keeping only the highest order terms in $c$, we find that the problematic off-diagonal terms appearing in Equation \eqref{eq:PQG-actionNewton} disappear. In other words, we show that the dynamics of Equation \eqref{eq:PQG-actionNewton} defines CP dynamics on the subset of states defined by the Newtonian limit. We leave it as a question for further work whether the CQ constraints would be preserved in the more general case, and in particular, if the dynamics of Equation~\eqref{eq:PQG-actionNewton} lead to stable dynamics which preserves the Newtonian limit once higher order terms are $c$ are considered.


Consider the full diffeomorphism invariant theory of CQ general relativity, which, when coupled to a quantum mass density, has a path integral of the form:
\begin{equation}
\label{eq: GRPathMain}
    \varrho(g_f,\phi^+_f,\phi^-_f,t_f)= \mathcal{N}\int \mathcal{D} g\mathcal{D}\phi^+\mathcal{D}\phi^-  \; e^{\mathcal{I}_{CQ}[g,\phi^+,\phi^-_,t_i,t_f]}  \varrho(g_i,\phi^+_i,\phi^-_i,t_i) ,
\end{equation}
where $\mathcal{N}$ is a normalisation factor and the action takes the form of:
\begin{equation}
\label{eq:PQG-actionNewton}
\begin{split}
    \mathcal{I}_{CQ}[g,\phi^+,\phi^-,t_i,t_f] &=\int_{t_i}^{t_f} dtd\vec{x}\, \bigg[ i\big(\mathcal{L}_{Q}[g,\phi^+]-\mathcal{L}_{Q}[g,\phi^-]\big)\\&\quad -\frac{\text{Det}[-g]}{8}\big(T^{\mu\nu}[\phi^+] - T^{\mu \nu}[\phi^-]\big) D_{0,\mu \nu\rho \sigma}[g]\big(T^{\rho\sigma}[\phi^+] - T^{\rho \sigma}[\phi^-]\big) \\
    & \quad - \frac{\text{Det}[-g] c^8}{128\pi^2 G_N^2 } \left( G^{\mu \nu} - \frac{8 \pi G}{c^4}\bar{T}^{\mu\nu}[\phi^+,\phi^-]  \right) D_{0, \mu \nu \rho \sigma}[g] \left( G^{\rho \sigma} - \frac{8 \pi G}{c^4}\bar{T}^{\rho\sigma}[\phi^+,\phi^-]  \right)
    \bigg],
    \end{split}
\end{equation}
where $\mathcal{L}_{Q}$ is the quantum Lagrangian density including the appropriate metric factors, $\bar{T}[\phi^+,\phi^-]$ is the average of the left and right branches of the stress-energy tensor and  we have taken $D_0, D_2$ to saturate the decoherence-diffusion trade-off \eqref{eq: decdiffsat} such that both the decoherence and diffusion coefficients are written in terms of $D_0$. Here, the bra and ket fields $\phi^\pm$ can be any quantum fields, but we shall consider pressureless dust $m^\pm(x)$ as a special case.

We now take the couplings to be ultra-local so that we can write them in terms of the generalised deWitt metric \cite{dewitt_1967,giulini_kiefer_1994}:
    \begin{align}
        D_{0,\mu \nu \rho \sigma}=\frac{1}{2}\frac{D_0}{\sqrt{-g}}\big(\,g_{\mu\rho} g_{\nu \sigma}+g_{\mu \sigma} g_{\nu \rho}-2\beta g_{\mu\nu} g_{\rho \sigma}\big).
    \end{align}
To obtain the Newtonian limit, we write the path integral in an ADM formalism, described by summing over all paths of the Lapse, Shift, and spatial metric ($N, N^i, \gamma_{ij}$) and inserting the choice of matter field as a pressureless dust distribution.
We then consider the action as a functional of the variables appearing in the ADM decomposition $\mathcal{I}_{CQ}[N,\vec{N},\gamma_{ij},m^+,m^-]$.

The Newtonian limit can then be understood as a gauge fixing of the complete theory, computing the transition amplitudes between the CQ states defined on hypersurfaces $\Sigma_t$:
\begin{equation}
\begin{split}
    \varrho(\gamma_f,m^+_f,m^-_f,t_f)&= \int \mathcal{D} \gamma \mathcal{D}N \mathcal{D} \vec{N} 
    \mathcal{D}m^+\mathcal{D}m^-  \;\delta\left(\gamma_{ij} - \left(1- \frac{2 \Phi}{c^2} \right)\delta_{ij}\right) \delta\left( N- \left(1+ \frac{\Phi}{c^2}\right)\right) \\
    &\quad \times \delta\left( N^i- \frac{n^i}{c^3}\right) e^{\mathcal{I}_{CQ}[N,\vec{N},\gamma_{ij},m^+,m^-,t_i,t_f]}  \varrho(\gamma_i,m^+_i,m^-_i,t_i)   .
\end{split}
\end{equation}

Performing the delta functional integrals, we impose the Newtonian gauge. In particular, we have $g_{00} = -(cN)^2 \approx -(c^2+ 2\Phi)$, whilst $g_{ij} \approx \left(1-\frac{2 \Phi}{c^2}\right)$ and $g_{0i} = \frac{n^i}{c^3}$. The components of the Einstein tensor are calculated as 
  \begin{align}
        &G_{00}=-2\nabla^2\Phi,\\
        &G_{0i}=-\frac{2}{c^5}\partial_0\partial_i\Phi+  \frac{1}{2 c^5}\nabla^2n_i,\\
        &G_{ij}=- \frac{2}{c^4}\partial_t \partial_t \Phi .
    \end{align}
Similarly, noting that $\det(-g) \approx c^2$, we see that due to the powers of $c$, the de-Witt metric is dominated by its $0000$ component, which to leading order is given by
\begin{equation}
    D_{0,0000}=D_0 c^3( 1-\beta).
\end{equation}

Keeping only terms leading order in $c$, the path integral action in Equation \eqref{eq:PQG-actionNewton} is dominated by terms only involving $D_{0000}$ and leads to the Newtonian path integral: 
    \begin{equation}
    \label{eq: PathNewt}
        \varrho(\Phi_f,m^+_f,m^-_f,t_f)= \mathcal{N}\int \mathcal{D} \Phi\mathcal{D}m^+\mathcal{D}m^-  \; e^{\mathcal{I}_{CQ}[\Phi,m^+,m^-,t_i,t_f]}  \varrho(\Phi_i,m^+_i,m^-_i,t_i), 
    \end{equation}
with CQ action given by:
\begin{equation}
\label{eq: PQG-actionNewton2}
\begin{split}
    \mathcal{I}_{CQ}[\Phi,m^+,m^-,t_i,t_f] = \int_{t_i}^{t_f} dtd\vec{x}\, &\bigg[i\big(\mathcal{L}_{Q}[m^+]-\mathcal{V}_I[\Phi,m^+]-\mathcal{L}_{Q}[m^-]+\mathcal{V}_I[\Phi,m^-]\big)\\& -\frac{\tilde{D}_0}{2}\big( m^+(x) - m^-(x)\big)^2  - 2\tilde{D}_0 \left( \frac{\nabla^2 \Phi}{4 \pi G } - \bar{m}(x)\right)^2 \bigg].
\end{split}
\end{equation}
were $\bar{m}(x)=\frac{1}{2}\big(m^+(x)+m^-(x)\big)$, we have redefined $\tilde{D}_0 = \frac{c^5 D_0}{4}(1-\beta)$, $\mathcal{L}_Q[m^\pm]$ is the matter Lagrangian in flat spacetime and $\mathcal{V}_I[\Phi,m^\pm]$ is the interaction potential coming from the expansion of $\sqrt{-g}\approx c-2\Phi$ in the matter Lagrangian for curved spacetime. Sources coupled to the classical or quantum degrees of freedom could be added if needed.

We have arrived at the final form of the Newtonian CQ path integral. Since it gives the state of the gravitational field for any quantum state of matter, it can be thought of as the constraint equation, consistent with the proposal in \cite{UCLPISHORT}. This is one of the main results of this paper. Equation \eqref{eq: PQG-actionNewton2} describes an integral over paths of the classical Newtonian potentials and a doubled path integral over the quantum mass eigenstates $m^{\pm}$, which occur because the integral is a density matrix path integral over both bra and ket branches.
Hence, provided $\beta \leq 1$, we find that the dynamics of the $c\to \infty$ limit of the full theory of Equation \eqref{eq:PQG-actionNewton} gives rise to complete positive evolution, which describes a randomly sourced Poisson's equation with associated decoherence into mass eigenstates of the quantum state. This justifies from a top-down approach the Newtonian limit that will be derived in Equation \eqref{eq: constraintfinal} and gives rise to the hope that the theory \cite{UCLPISHORT} has constraints that are preserved in time. However, we highlight that in this derivation we have arrived at the Newtonian limit by gauge fixing the full theory and neglecting all the terms of higher order in $c$. Since we have neglected terms of higher order in $c$, we have implicitly got rid of the potentially positivity-violating terms of the full path integral \eqref{eq:PQG-actionNewton}. The terms involving $D_{000i}, D_{0i0j}$ still arise, but they are higher order in $c$. Therefore, we have shown that the dynamics of Equation \eqref{eq:PQG-actionNewton} defines CP dynamics on the \emph{subset} of states defined by the Newtonian limit but we have not shown that the dynamics is consistent away from this limit. For example, the evolution could be unstable for finite c, but we leave this as a question for future research. Nevertheless, a more general treatment is essential since we have not shown that the complete theory necessarily preserves the form of the Newtonian limit. A possible outcome might be that including higher-order terms in the calculation causes the dynamics to somehow deviate from the correct limit. The clearest example of this hypothetical behaviour can be seen by considering that, in deriving the Newtonian limit, we have assumed that Poisson's equation holds on average at any scale.  

In the next section, we will introduce the master equation formalism for CQ dynamics and its unravelling. These will then be used to derive the Newtonian limit in this formalism. The dynamics will match the behaviour derived in Equations~\eqref{eq: GRPathMain} and \eqref{eq:PQG-actionNewton}. Readers content with the path integral formulation could skip to the Discussion section.

\section{Master equation formalism and CQ Unravelling}
\label{sec: MasterEQ}

In this section, we review the master equation and the unravelling formalism and comment on the relationship between the master equation and the decoherence-diffusion trade-off derived in \cite{UCLdec_Vs_diff}. 
One can, in theory, also go from the master equation to the corresponding path integral as described in \cite{UCLPILONG}. For the interested reader, we do this in Appendix~\ref{app: PInewt}.
Given the hybrid CQ state defined in Section~\ref{sec: CQdyn} one can write the most general dynamics of these states such that they retain positivity, preserve the statistical interpretation of the density matrix and give rise to positive probabilities when acting on half-entangled states. This implies that this dynamics must be linear, completely positive (CP) and probability-preserving. With the additional assumption of time-locality, the dynamics can be written in the form of a CQ master equation \cite{UCLpost_quantum}:

\begin{equation}
\frac{\partial \cqstate(z,t)}{\partial t}  =  \int dz' \ W^{\mu \nu}(z|z') L_{\mu} \cqstate(z') L_{\nu}^{\dag} - \frac{1}{2}W^{\mu \nu}(z) \{ L_{\nu}^{\dag} L_{\mu}, \cqstate \}_+,
\label{eq: cqdyngen}
\end{equation}
where $\{, \}_+$ is the anti-commutator, $L_\mu$ are an arbitrary set of operators on the Hilbert space known as \textit{Lindblad operators}. Preservation of normalization under the trace and $\int dz$ defines
\begin{equation}
W^{\mu  \nu}(z) = \int \mathrm{d} z' W^{\mu \nu}(z'| z).
\end{equation}
Introducing a basis $L_{\mu} = (\mathbb{I}, L_{\alpha})$ of Lindblad operators, the condition for Equation \eqref{eq: cqdyngen} to describe completely positive dynamics at all times is that the matrix
\begin{equation} \label{eq: block1}
\begin{bmatrix}
    \delta(z,z') + \delta t W^{00}(z|z')      & \delta t W^{0\beta}(z|z') \\
    \delta t W^{\alpha 0}(z|z')    & \delta t W^{\alpha \beta} (z|z')  \\
\end{bmatrix} 
\end{equation}
be a completely positive matrix kernel in $(z,z')$ \cite{UCLdec_Vs_diff}.
The CQ master equation is therefore seen to be a natural generalization of the Lindblad equation and classical rate equation in the case of classical-quantum coupling.  For notation simplicity, we shall take the dynamics to be autonomous, which we take to mean that $W^{\mu \nu}(z|z')$ are time-independent. 

The master equation \eqref{eq: cqdyngen} can be expanded in terms of the moments $D_n$ of the transition amplitude $W^{\mu \nu}(z|z')$ \cite{UCLpost_quantum,UCLdec_Vs_diff, UCLPILONG}, 
\begin{equation}
  D^{\mu \nu}_{n, i_1 \dots i_n}(z'):= \frac{1}{n!}\int dz W^{\mu \nu}(z|z')(z-z')_{i_1} \dots (z-z')_{i_n}.
\end{equation}
It is important to note that the moments $D_n$ are not independent since they must define complete positive dynamics and inherit the complete positivity conditions through 
the moments' expansion of the transition amplitudes. In \cite{UCLdec_Vs_diff}, it was shown that in order for the dynamics to be completely positive, then
\begin{equation}\label{eq: generalTradeOffCouplingConstants}
     \ D_1^{br} D_0^{-1} D_1^{ br \dag} \preceq  D_2
\end{equation}
holds for the matrix whose elements are the couplings $D_2 =D^{\mu \nu}_{2,ij} ,D_1^{br}= D^{\alpha \mu}_{1,i}, D_0 =D_0^{\alpha \beta}$. Moreover $D_0, D_2$ are required to be positive semi-definite and $ (\mathbb{I}- D_0 D_0^{-1})D_1^{br} =0$, which tells us that $D_0$ cannot vanish if there is non-zero back-reaction. Equation \eqref{eq: generalTradeOffCouplingConstants} tells us that whenever there is back-reaction of a quantum system on a classical one, we \textit{necessarily} have decoherence on the quantum system, as well as diffusion in the classical system, by an amount lower bounded by the coherence time. The trade-off in Equation \eqref{eq: generalTradeOffCouplingConstants} must hold for \textit{all} Markovian classical-quantum dynamics. We refer to the trade-off in Equation \eqref{eq: generalTradeOffCouplingConstants} as the \textit{decoherence-diffusion trade-off}, though strictly speaking, it is a trade-off between the diffusion $D_2$ and Lindbladian coupling $D_0$ entering into the master equation. A special case of this can be found in the condition for complete positivity of the constant force master equation of \cite{diosi1995quantum}.

It was further shown in \cite{UCLpawula} that such master equations can be split into two classes, those which have continuous trajectories in the classical space, first discovered in \cite{diosi1995quantum}, and those with finite-sized jumps \cite{blanchard1995event}. The most general form of the continuous master equation is given by: 

\begin{align} \label{eq: CQcontinuous}\nonumber
\frac{\partial \cqstate(z,t)}{\partial t} & =  \sum_{n=1}^{n=2}\frac{(-1)^n}{n!}\left(\frac{\partial^n }{\partial z_{i_1} \dots \partial z_{i_n} }\right) \left( D^{00}_{n, i_1 \dots i_n} \cqstate(z,t) \right) - \frac{\partial }{\partial z_{i}} \left( D^{0\alpha}_{1, i} \cqstate(z,t) L_{\alpha}^{\dag} \right)  - \frac{\partial }{\partial z_{i}} \left( D^{\alpha 0 }_{1, i} L_{\alpha} \cqstate(z,t) \right)   \\ 
& -i[H(z), \cqstate(z,t)] + D_0^{\alpha \beta}(z) L_{\alpha} \cqstate(z) L_{\beta}^{\dag} - \frac{1}{2} D_0^{\alpha \beta} \{ L_{\beta}^{\dag} L_{\alpha}, \cqstate(z) \}_+  ,
\end{align}

where the decoherence-diffusion trade-off reads $D_{2}^{00} \succeq D_1 D_0^{-1} D_1^{\dag}$ and $(\mathbb{I}- D_0 D_0^{-1})D_1 =0$.
This expansion might appear cumbersome but can be readily interpreted as a sum of distinct effects. The first term represents the classical evolution of the system. It is composed of derivatives with respect to the classical degrees of freedom, where the first-order terms represent the Liouville evolution and the second-order terms, with coefficients $D_2^{00}(z)$, represent the diffusion in phase space. The second and third terms encode the back-reaction of the quantum degrees of freedom on the classical degrees of freedom, given by the coupling of derivatives with the Lindblad operators. These terms' coefficient is $D_1(z)$. The first term of the second line is the usual unitary quantum evolution, while the last two terms, with coefficient $D_0(z)$, are responsible for the decoherence of the quantum system. When the drift is generated by a $CQ$ Hamiltonian, the back-reaction described by the $D_1$ term takes the form of the \textit{Alexandrov-Gerasimenko bracket} and the decoherence-diffusion trade-off originating from the requirement of positivity becomes $4D_2\succeq D_0^{-1}$ \cite{UCLdec_Vs_diff}, which, when saturated, takes the form of Equation~\eqref{eq: decdiffsat}, ensuring the positivity of the CQ path-integral.

Since gravity is a field theory, we need to use the field-theoretic version of the moment expansion which was studied in \cite{UCLpost_quantum, UCLconstraints,UCLdec_Vs_diff}. In the field-theoretic case, the Lindblad operators can have spatial dependence, and the field-theoretic master equation follows by replacing discrete sums with integrals over space
and replacing standard derivatives with functional derivatives. In other words, the spatial coordinate $x$ acts like an index of the Lindblad operators and the matrices $D_n$. We will be explicit with this in section \ref{sec: CQnewtlim} and refer the reader to \cite{UCLpost_quantum, UCLdec_Vs_diff} for a more detailed discussion of the master equation derivation in this case. 

In the field-theoretic case, one finds the same trade-off between coupling constants in Equation \eqref{eq: generalTradeOffCouplingConstants}, but the moments are now matrix kernels representing diffusion and decoherence. In order for the dynamics to be completely positive $D_0(x,y), D_2(x,y)$ must also be positive kernels, where a positive kernel $f(x,y)$ is a kernel such that $\int \ dx dy a^*(x)f(x,y) a(y) \geq 0$ for any function $a(x)$. 
This is equivalent to a trade-off between coupling constants in Equation \eqref{eq: generalTradeOffCouplingConstants} when viewing it as a matrix-kernel equation.

When working with an open quantum system that is evolving according to the GKSL (Lindblad) equation, it is often convenient to recast the dynamics in an unravelled form, where one follows the evolution of a single pure quantum state instead of following the entire density matrix. The state evolves stochastically in the Hilbert space, and the evolution of the quantum system can then be recovered by averaging over all paths. The advantages of this approach are multiple. Firstly, it is easily implementable in a computer simulation; the evolution of a single pure state, especially for larger systems, has lower computational complexity than the entire density matrix. Secondly, the unravelling approach is intuitively easier to grasp and offers a different perspective. The evolution of the system can be thought of as generated by continuous, deterministic dynamics accompanied by stochastic jumps of the wave function occurring stochastically whenever an interaction with the environment manifests. 

Much like the GKSL equation, the CQ master equation can be unravelled to study the stochastic evolution of the quantum and classical degrees of freedom of a pure hybrid CQ state.


In this work, the role of classical degrees of freedom will be played by the Newtonian gravitational scalar field and its conjugate momentum, while the density matrix will describe the quantum matter degrees of freedom. The dynamics of semi-classical hybrid gravity can then be visualized as physical stochastic paths through configurations of the fields, where measurements of the quantum mass act as sources for the classical gravitational degrees of freedom. For more information on the unravelling of CQ equations, we point the reader to Appendix \ref{app: unravelling} and \cite{tilloy2016sourcing,UCLunrav,UCLpawula, UCLdec_Vs_diff, UCLhealing}.
In the next section, we will discuss the master equation describing the Newtonian gravitational field coupled with the quantum matter degrees of freedom, such that the Newtonian interaction is reproduced on average.



\section{Hamiltonian CQ dynamics reproducing gravity in the Newtonian limit}
\label{sec: CQnewtlim}

Having discussed in detail the Newtonian limit of GR, we are now in a position to discuss how to write down classical-quantum theories that give rise to the Newtonian interaction on average. Before discussing the specifics of continuous and discrete master equations, we shall outline the general procedure and assumptions.

\begin{assumption}
We assume that the CQ dynamics takes the form of Equation \eqref{eq: cqdyngen}
\end{assumption}
Since Equation \eqref{eq: cqdyngen} is the most general form of Markovian, classical-quantum evolution, we expect this assumption to hold if we are treating this as a fundamental theory. If on the other hand, it is an effective theory, the Markovian assumption may break down. We comment on the differences between a fundamental and effective theory in section \ref{sec: discussion} (see also the appendices of \cite{UCLhealing}).
\begin{assumption}
 In the weak field $c\to \infty$, the appropriate gravitational degrees of freedom are the perturbations of the metric in the form of Equation \eqref{eq: approxMetricSimple}. 
\end{assumption}
In particular, the leading order contribution which governs the geodesics of test particles is described by $h_{00}$. This is a bottom-up approach in the sense that we reduce the degrees of freedom in the action, and then construct the CQ theory.
\begin{assumption}
We assume that the purely classical part of the evolution is generated by the reduced Hamiltonian \eqref{eq: reducedHamEquals}, that the interaction between classical and quantum degrees of freedom is Hamiltonian and that it is governed by the reduced interaction Hamiltonian in Equation \eqref{eq: intHam}, where the constraints $\pi_{\Phi}, \pi_i \approx 0 $ should also be imposed. 
\end{assumption}
Specifically, we require that the first moment $D_1$ is picked to reproduce the Newtonian back-reaction on average: 
\begin{equation}\label{eq: newtexpectation}
   \Tr{}{\{ H^{I}, \cqstate \} } =  \int d^3 x \ \Tr{}{ \hat{m}(x) \frac{\delta \rho}{\delta \pi_{\Phi}(x)} }, 
\end{equation}
so that the dynamics are Hamiltonian on average. While this might appear to be a mild assumption, it does assume that the coupling strength and gravity itself is either unmodified at arbitrarily short-distances, or at least that the short distance behaviour does not affect physics in the weak field regime.

 Since the back-reaction of the quantum system on the classical system is associated with $T_{00}$, we expect the CQ momentum constraint to be unchanged from its classical counterpart, as it is not associated with any back-reaction. This was also found to be the case in a study of CQ gravitational constraints \cite{UCLconstraints}, and in the scalar gravity theory we consider in \cite{UCLNordstrom}.

\begin{assumption}
In this work, we will take the coefficients $D_n$ entering the master equation to be minimally coupled, by which we mean they depend only on the Newtonian potential $\Phi$, $D_n(\Phi)$ and not their conjugate momenta $\pi_{\Phi}$. 
\end{assumption}
This assumption is motivated by the fact in Einstein's gravity, the mass density couples to the Newtonian potential and not its conjugate momenta, and we are imposing the constraint that $\pi_{\Phi} \approx 0$. Nonetheless, one could generalize the master equations to the non-minimally coupled case by considering couplings $D_n(\Phi) \to D_n(\Phi, \pi_{\Phi})$ in all of the equations. 

We now consider the dynamics consistent with assumptions 1-4. In the main body, we only discuss the weak field limit for continuous master equations since, in this case, we can be more thorough and explicit. Then, we will use the unravelling of the weak field limit to impose the Newtonian constraints and obtain our main result as a set of coupled stochastic differential equations describing the Newtonian CQ dynamics.

\subsection{Continuous gravitational back-reaction }

In \cite{UCLpawula}, it was shown that there are two classes of CQ master equations. One of the forms includes finite-sized jumps in the classical degrees of freedom due to the backreaction of the quantum part of the system, while in the other the evolution remains continuous. The most general form of the CQ continuous master equation was then explicitly given as Equation \eqref{eq: CQcontinuous}. 
When the back-reaction is continuous, specifying that the first moment on average satisfies Equation~\eqref{eq: newtexpectation} is enough to fix the terms of Equation~\eqref{eq: CQcontinuous} which correspond to the continuous back-reaction of the classical and quantum degrees of freedom onto each other. As we discuss below in more detail, this is only the continuous part of the backreaction, since the stochastic nature imposed by the Newtonian constraints on $n^i$ will introduce further jumping backreaction terms. The continuous backreaction is composed of the decoherence and diffusion effects described by:

\begin{equation}\label{eq: contbackreaction}
\begin{split}
&  \frac{1}{2}\int d^3x \big( \{\mathcal{H}^I(x),\cqstate\}-\{\cqstate,\mathcal{H}^I(x)\} \big)
+\int d^3x d^3y\,  
D_{2}(\Phi; x,y)\frac{\delta^2 \cqstate}{\delta \pi_\Phi(x)\pi_\Phi(y)}  \\
&\quad\quad\quad + \frac{1}{2}\int d^3 x d^3y\, D_0(\Phi;x,y)\left( [ \hat{m}(x), [ \cqstate, \hat{m}(y) ] ] \right),
\end{split}
\end{equation}
where $\pi_{\Phi}$, and $D_0(\Phi;x,y), D_2(\Phi;x,y)$ are positive semi-definite kernels\footnote{Recall a positive semi-definite kernel $f(x,y)$ is a kernel such that $\int dx dy a^*(x)f(x,y) a(y) \geq 0 $ for any function $a(x)$.},  and $\mathcal{H^{I}}=\Phi(x)\hat{m}(x)$ is the interaction Hamiltonian density. Here, $\hat{m}(x)$  is the quantum mass density operator. For notational simplicity, we shall often suppress the dependence of the couplings $D_0, D_2$ on the Newtonian potential and simply write $D_0(x,y), D_2(x,y)$. The Lindbladian term, characterized by $D_0$, and the diffusion term $D_2$ are required in order for the back-reaction to be completely positive, which can be seen from the decoherence diffusion trade-off \cite{UCLdec_Vs_diff} apparent in Equation \eqref{eq: CQcontinuous}. It is  possible to add extra diffusion or decoherence terms and still satisfy the conditions for the master equation to be completely positive, but here we only consider the minimal amount of decoherence and diffusion required. If one desires, the above equation may be written entirely in terms of Poisson brackets by defining a stochastic Hamiltonian as the integral of $\sqrt{D_2}$ over $\Phi$, and writing the diffusion term as a nested Poisson bracket with respect to it.


The entire master equation will also include terms associated with the pure Hamiltonian evolution of the quantum state, with the Hamiltonian given by  Equation~\eqref{eq: reducedHamEquals}, 
    \begin{equation}
        H^{(gr)}+ H^{I} 
        =\int d^3x \left[ \frac{(\nabla \Phi)^2}{8\pi G}+  \hat{m}\Phi  -\frac{1}{12} \pi_{\Phi} \partial_i n^i +  \lambda^i \pi_i \right],
    \end{equation}
where the mass density $m(x)$ has been replaced by the operator $\hat{m}(x)$. This Hamiltonian contains $\Phi$ and $n^i$ which will become stochastic degrees of freedom in order to impose the $\pi_\Phi\approx 0$ constraint. As a result, care must be taken, since this will result in the master equation containing additional Fokker-Plank and jump terms of the form
    \begin{equation}
        \alpha(\Phi,n^i)\frac{\delta^{k+l}\varrho}{\delta^k\Phi\delta^l\pi_\Phi} \quad k+l\geq 2
    \end{equation}
associated with the correlations between $\Phi$ and $n^i$. Choosing $n^i$ stochastically will back-react on $\Phi$, giving a master equation with infinite terms that enforce the constraint by forcing the shift to have the necessary correlation with the stochastic gravitational field.

Since the only degree of freedom in Equation \eqref{eq: newtlimitdyn} associated with the matter back-reaction is $\pi_{\Phi}$, up to these nuances, the choice of the possible master equation is therefore fully constrained up to the functional choices of the couplings $D_0(\Phi;x,y), D_2(\Phi;x,y)$, which are themselves constrained to satisfy the decoherence diffusion trade-off: 
\begin{equation}\label{eq: decdiffcont}
    4 D_2 \succeq D_0^{-1}.
\end{equation}
Equation \eqref{eq: decdiffcont} is to be understood as a matrix kernel equation:
\begin{equation}\label{eq: conttradeoffcomponents}
    \int dx dy \ a(x)^* \big[ 4 D_2(\Phi;x,y) -D_0^{-1}(\Phi; x,y)\big] a(y) \geq 0,
\end{equation}
which must hold for an arbitrary function $a(x)$. In Equation \eqref{eq: conttradeoffcomponents}, $D_0^{-1}(\Phi; x,y)$ is the generalized kernel inverse of the diffusion coupling $D_0(\Phi; x,y)$~\cite{UCLdec_Vs_diff} which is only required to be a positive semi-definite kernel. We give example kernels that satisfy the decoherence diffusion trade-off in Table \ref{table: kernels}. The decoherence diffusion condition in Equation \eqref{eq: decdiffcont} can be used to experimentally constrain fundamental theories with a classical gravitational field~\cite{UCLdec_Vs_diff}. Before discussing the experimental bounds on the dynamics described by Equation \eqref{eq: contbackreaction}, we must first impose the Newtonian constraint $\pi_{\Phi} \approx 0 $.

\subsection{Newtonian unravelling of the master equation}\label{sec: imposingconstraint}

To arrive at the classical-quantum version of Poisson's equation, we must impose the constraint $\pi_{\Phi} \approx 0$ according to the Hamiltonian in Equation \eqref{eq: reducedHamEquals}.
In classical Hamiltonian dynamics, typically, one imposes constraints on initial data. One then enforces the preservation of the constraints by the dynamics via the Dirac procedure \cite{diracbook}. Because in Hamiltonian dynamics, there is an isomorphism between initial physical data and physical solutions to the dynamics, this procedure is equivalent to imposing the constraint on solutions to the dynamics since, by construction, the dynamics never leave the constraint surface.

In the non-deterministic case, where the dynamics contain randomness, there is no isomorphism between initial data and the solution space of the dynamics. Instead, one builds a probability distribution over the possible trajectories of the initial data according to the dynamics and imposes constraints on the trajectories.\footnote{This is conceptually very similar to what is done in quantum theory when constraints are imposed via a path integral approach, where one associate to each path a measure given by the action, then selects only the paths which satisfy the constraint. Take for example, a Hamiltonian with $H(q,p) = H_0(q,p) + \lambda C(q,p)$. The phase space partition function for the theory is represented by $\mathcal{Z} = \int \mathcal{D} q \mathcal{D} p \mathcal{D}\lambda e^{\frac{i}{\hbar} \int dt [\dot{q} p - H(q,p) - \lambda C(q,p)]}$. Since the Hamiltonian is linear in $\lambda$, the path integral over the Lagrange multiplier in $\lambda$ enforces a delta function over $\delta(C(q,p))$ so that the partition function reads  $\mathcal{Z} = \int \mathcal{D} q \mathcal{D} p \delta(C(q,p)) e^{\frac{i}{\hbar} \int dt [\dot{q} p - H(q,p) ]}$ which can be interpreted as summing over all paths with weight $e^{\frac{i}{\hbar} \int dt [\dot{q} p - H(q,p) ]}$ and then selecting only those that satisfy the constraint $C(q(t),p(t))=0$.}  As such, when we implement the constraint in the classical-quantum case, we have to ensure that the dynamics remain CP.

Given that we are imposing the constraint $\pi_{\Phi} \approx 0$ on the level of trajectories, it is more convenient to go to an unravelling picture, which enables us to discuss explicitly classical-quantum trajectories which satisfy the constraint. 
We consider the unravelling picture of CQ dynamics to be a clear way of presenting these results of the paper, also because it allows for an ontological interpretation of the trajectories and for the ease associated with simulating its time development with a computer. It is also natural in the relativistic case since the trajectory of 3-geometries is the space-time geometry.

The unravelling of the weak field master equation with continuous backreaction given by Equation~\eqref{eq: contbackreaction}  is derived by substituting the Hamiltonian drift terms in Equation~\eqref{eq: unravel_general} which is the general expression for the unravelling of the continuous master equation derived in \cite{UCLhealing}. Recalling that the mass density operator $\hat{m}$ is Hermitian, this results in the following coupled It\^o stochastic differential equations:

\begin{align}
    \label{eq: newtUnrav}
    & d \Phi_t =  - \frac{1}{12} \partial_i n^i  dt, \\
    & d \pi_{\Phi\,t} = \frac{\nabla^2 \Phi_t}{4 \pi G }dt- \langle \hat{m}(x) \rangle dt - \int d^3y\, \sigma(\Phi_t;x,y) d W_t(y),\\
    & d \rho_t = -i[\hat{H}_m+\hat{H}^I,\rho_t] dt + \frac{1}{2}\int d^3x\, d^3 y\, D_0(\Phi_t; x,y) [ \hat{m}(x), [ \rho_t, \hat{m}(y) ] ]  dt \nonumber \\
    & \quad \quad \quad+ \frac{1}{2}\int d^3x\,d^3y\, \sigma^{-1}(\Phi_t;x,y) \big(\hat{m}(x)\rho_t+\rho_t\, \hat{m}(x)-2\rho_t \langle \hat{m}(x) \rangle\big) d W_t(y) \label{eq:statechange}
\end{align}
where $\hat{H}_m$ is the matter Hamiltonian, $\hat{H}^I=\int d^3x\, \hat{m}(x) \Phi(x)$ is the interaction Hamiltonian, $\langle\cdot\rangle$ is the usual expectation value, $\rho(t)$ is the normalized quantum state and $W_t(x)$ is a Wiener process in spacetime satisfying:

\begin{equation}
\label{eq: Wiener}
    \mathbb{E}[dW_t(x)]=0, \ \ \mathbb{E}[dW_t(x) dW_t(y)] =  \delta(x,y) dt.
\end{equation}
In Equation \eqref{eq: newtUnrav}, $\sigma$ and its generalized inverse $\sigma^{-1}$ are related to the diffusion coefficient $D_2$ appearing in Equation \eqref{eq: contbackreaction} via:

\begin{equation}
\label{eq: diffcoeff}
   D_2(\Phi; x,y) =  \int dw  \ \sigma(\Phi;x,w)\sigma(\Phi; y,w) .
\end{equation}
One can verify that this unravelling is equivalent to the CQ master equation with continuous backreaction given by Equation~\eqref{eq: contbackreaction} by using it to compute the evolution of the CQ state defined via:     \begin{equation}
 \varrho(\Phi, \pi_{\Phi},t) = \mathbb{E}[\delta(\Phi_t -\Phi)\delta(\pi_{\Phi\,t} -\pi_{\Phi}) \rho_t].
 \end{equation}
We show this explicitly in Appendix~\ref{app: unravelling}, which is enough to verify that this is the correct unravelling for the master equation since the unravelling is unique \cite{UCLhealing}. The classical theory with  Hamiltonian written in Equation~\eqref{eq: reducedHamEquals} which we make stochastic in Section~\ref{sec: Newtlim}  is equivalent to the dynamics for $\Phi$ in Equation~\eqref{eq: newtUnrav} once we have imposed the $\pi_\Phi\approx 0$ constraint.
The clear difference is that in the CQ case, the noise process is not added manually, but emerges due to positivity requirements after the direct coupling of the quantum matter degrees of freedom with the classical Newtonian potential. 
Hence, it is the back-reaction in $\pi_{\Phi}$ that turns it into a stochastic process. Moreover, the noise now is correlated with the quantum state, and the evolution of the quantum state itself involves decoherence due to the backreaction on the classical Newtonian potential. Indeed, the evolution of the quantum state is equivalent to that of a state undergoing continuous measurement of its mass, although the quantum state is pure conditioned on the evolution of the Newtonian potential. It is only after tracing out the gravitational field that the decoherence is made manifest. 
We refer the reader interested in a detailed discussion to \cite{UCLhealing}. In particular, this defines a linear master equation.


In order to arrive at the full Newtonian limit, the final step consists of imposing the constraint $\pi_{\Phi} \approx 0$. This can be done directly on the classical quantum evolution of Equation \eqref{eq: newtUnrav} or at the path integral level, through the use of delta functionals.
We present here the former way, but the latter procedure is presented in Appendix \ref{app: PInewt} where, after constructing the path integral for the reduced gravitational degrees of freedom, we impose the Newtonian constraints at the level of trajectories.





In order to impose the constraint $\pi_{\Phi} \approx 0$ on Equation \eqref{eq: newtUnrav}, one must choose $n_i$ stochastically such that $\dot{\Phi}\approx\dot{\pi}_\Phi\approx 0$, where the equality is the weak equality in the Dirac sense. Doing so turns $\Phi_t$ into a white noise variable with values given by the solution of Equation~\eqref{eq: constraintfinal}. However, naively replacing all occurrences of $\Phi_t$ with its solution in terms of an It\^o white noise variable, in particular, that which appears in $\hat{H}^I$, does not lead one to a CPTP dynamics. Before the constraints are imposed, the dynamics of $\Phi_t$ are continuous, and thus any back reaction from the quantum matter on $\Phi_t$ only returns to affect the quantum matter degrees of freedom at later times. To ensure that this time-ordering is maintained even in the limit that $\Phi_t$ no longer evolves continuously, one must be careful to ensure the action of $\hat{H}^I$ occurs after the other stochastic terms (for excellent further discussion on this issue of time-ordering, we refer the reader to \cite{tilloy2016sourcing}, and \cite{wiseman93feedback.70.548,diosi94comment}). One possible way to ensure this is to first write the unravelling of the density matrix in the Stratonovich form and then impose the constraint that turns $\Phi_t$, and hence $\hat{H}^I$, into white noise. This allows us to correctly rewrite the unravelling such that when converting back into the It\^o formalism, we pick up an extra decoherence term given by the backreaction of the stochastic gravitational field and allows us to get rid of the non-linear evolution term picked up from the solution of the noise Poisson equation. This then gives the final form of unravelling in the Newtonian limit:


\begin{align}
\label{eq: constraintfinal}
& \frac{\nabla^2 \Phi_t}{4 \pi G } =  \langle \hat{m}(x) \rangle + \int d^3y\, \sigma(\Phi_t;x,y) \xi_t(y) ,\\
& d \rho_t = -i\left[\hat{H}_m+\hat{V}_{m},\rho_t\right]dt-i\int d^3x\,d^3y \,d^3y'\left[-G\,\frac{\hat{m}(x)\sigma(\Phi_t,y,y')}{|x-y|},\rho_t\right]dW_t(y')  \nonumber \\
&\quad \quad \quad+\frac{1}{2}\int d^3x\, d^3 y\, D_0(\Phi_t; x,y)\big( [ \hat{m}(x), [ \rho_t, \hat{m}(y) ] ] \big) \, dt \nonumber\\
&\quad \quad \quad+\frac{1}{2}\int d^3x\, d^3y\, d^3 y'\,[\hat{\sigma}(\Phi_t;x,y),[\rho_t,\hat{\sigma}(\Phi_t;x,y')]]\,dt   \nonumber \\
\label{eq: constraintfinal_quantum}
& \quad \quad \quad+ \frac{1}{2}\int d^3x\, d^3y\, \sigma^{-1}(\Phi_t;x,y) \big(\hat{m}(x)\rho_t+\rho_t\, \hat{m}(x)-2\rho_t \langle \hat{m}(x) \rangle\big)\, d W_t(y), 
\end{align}
where $\xi_t(x)=\frac{d W_t(x)}{dt}$ is the formal definition of white noise, and
\begin{equation}
\begin{split}
    & \hat{V}_{m}=-\frac{G}{2}\int d^3x\,d^3y\frac{\hat{m}(x)\hat{m}(y)}{|x-y|},\\
     &\hat{\sigma}(\Phi_t;x,y)=-G\int d^3\,y''\frac{\hat{m}(x)\sigma(\Phi_t,y,y'')}{|x-y|}.
    \end{split}
\end{equation}
These equations were first written down by Tilloy and Di\'osi in \cite{tilloy2016sourcing} and their derivation from a fundamental theory is a central result of our current work. In it, we notice how the Newtonian limit of CQ theories is described by a Newtonian potential diffusing around Poisson's equation by an amount defined by $D_2$, while simultaneously the density matrix decoheres into the mass eigenbasis by an amount determined by the Lindbladian coefficient $D_0$. 
In Equation \eqref{eq: constraintfinal}, the Newtonian potential changes in time due to the random noise process $W(x)$ and its evolution fixes the divergence of the now stochastic $ n^i$ which, in general, will be correlated with the noise process appearing in the evolution of the quantum state. The fact that $n^i$ is constrained not to vanish in order for the CQ theory to be consistent is another deviation from the standard Newtonian limit appearing in CQ gravity.

The details of the functional dependence of $\sigma$ and $D_0$ on $\Phi_t$ have been left unspecified. However, there are three notable classes of functional dependence that are worth highlighting. Firstly, they may not depend on $\Phi_t$ at all. In this case, the equations coincide with those of a continuous measurement of a quantum mass, where the measurement outcome is used to source the Newtonian potential, as given by equation (24) of \cite{tilloy2016sourcing}. Therefore, we have shown that such dynamics can be derived from classical-quantum theories of general relativity through a path integral approach, or through the unravelling of a completely positive master equation that agrees with the Newtonian limit on expectation. 
Secondly, one may consider $\sigma$ and $D_0$ to be dependent on a time integral of $\Phi_t$ i.e. $\int dt f(t)\Phi_t$ for an arbitrary function $f(t)$. If allowed, such theories would be non-Markovian, but still guaranteed to be CPTP. On the other hand, the final class of functional dependence is to allow a general Markovian functional of $\Phi_t$. This will generically lead to additional terms, as was observed with $H^I$ above, but these may not preserve the CPTP property of the dynamics. Exploring the details of these functional dependencies is an interesting question which we leave open for future work.

In Equations \eqref{eq: newtUnrav} and \eqref{eq: constraintfinal}  we have taken the drift to be local in $x$, while we allow for the possibility that the decoherence and diffusion terms could have some range. In this case, the interaction law between the classical and quantum systems is still local, but non-local correlations can be created \cite{OR-intrinsic}. Importantly, if the Lindbladian coupling $D_0(\Phi;x,y)$ has some range, then despite the fact the CQ interaction is local, the master equation can, in principle, generate entanglement between two spatially separated quantum systems via the Lindbladian coefficient though this effect is likely to be small.

One can constrain other Diffusion/Decoherence kernels via Equation \eqref{eq: constraintfinal} and the decoherence diffusion trade-off in Equation \eqref{eq: decdiffcont}. The Newtonian limit of CQ gravity predicts diffusion of the Newtonian potential by an amount depending on $D_2$. This can be upper bounded by precision mass experiments, which precisely measure the acceleration of particles. Conversely, coherence and heating experiments can be used to upper bound the inverse Lindbladian coefficient $D_0^{-1}$, which gives a lower bound on $D_2$ via the decoherence diffusion trade-off. Hence, when combined, it is possible to get an experimental squeeze on CQ theories. In \cite{UCLdec_Vs_diff} this was used to rule out ultra-local CQ theories, which are Equation's \eqref{eq: constraintfinal} and \eqref{eq: decdiffcont} when the couplings are taken to be delta functions $D_0(x,y), D_2(x,y) \sim \delta(x,y)$. 

Interestingly, when accounting for the stochasticity of the interaction Hamiltonian, the Newtonian limit we derive in Equation \eqref{eq: decdiffcont} contains a decoherence term proportional to $\sigma^2$, which gives 
bounds on decoherence due to constraints from anomalous heating.
For the case where the coupling constants are independent of the Newtonian potential, the effects of the additional decoherence term were considered in \cite{tilloy2017principle}. In particular,it was shown that the choice of kernel giving rise to minimal decoherence is the Di\'osi-Penrose kernel $D_0(x,y) =\frac{G}{|x-y|}$. The precise amount of decoherence depends on the cut-off of the system, and it was shown in \cite{tilloy2017principle} that theories with a cut-off below $10^{-15}m$ are inconsistent with experiments due to excess heating. This result calls for both an exploration of relativistic corrections to CQ theories, which we believe need to be considered at this scale, as well as experimental tests of 
gravity on smaller length scales.

\section{Discussion}
\label{sec: discussion}

In this work, we have considered, on general grounds, the weak field limit of classical-quantum theories of gravity, which give rise to linear, completely positive dynamics. 
The master equation we derived, is the weak field limit of the simplest realization of the relativistic theory in \cite{UCLpost_quantum}, while the path-integral we derive is the weak field limit of the manifestly covariant theory of \cite{UCLPISHORT}. Both yield the same theory in the weak field limit, as shown in Appendix~\ref{app: PInewt}. 
Our central result is that in the weak field limit, we arrive at Equation \eqref{eq: PQG-actionNewton2}.

We have here started from a fundamental, dynamical and relativistic theory, and it is worthwhile to compare the limit we arrive at, to previous 
models which have been proposed based on Newtonian gravity. 
An early model in which gravity is treated classically is the Schr\"odinger-Newton equation~\cite{pitaevskii1961vortex,gross1961structure,ruffini1969systems} which was also proposed as a model of gravitationally induced collapse of the wave-function \cite{diosi1987universal,penrose1998quantum}. However, because the dynamics is non-linear in the wave-function, it leads to instantaneous signalling \cite{gisin1989stochastic,gisin1990weinberg,polchinski1991weinberg} and a breakdown of the statistical mechanical interpretation of the density matrix. It is unrelated to the dynamics we have derived here which is linear.

The master-equation approach we use,
is more similar in spirit to that of  Di\'osi's \cite{diosi2011gravity}. 
Indeed, the Newtonian back-reaction in Equation \eqref{eq: contbackreaction} has some similarities with the one considered there, when the decoherence and diffusion kernels are chosen to be related to the Di\'osi-Penrose kernel (see Table \ref{table: kernels}). However, an important difference between our master equation and Di\'osi's is that it contains diffusion in $\pi_\Phi$, while here, we require $\pi_\Phi\approx 0$ as a constraint equation. This has to be the case here, because in the weak field limit the kinetic energy term in Equation \eqref{eq: totham}, $-\frac{2\pi Gc^2}{3}\pi_\psi^2$ is negative and the theory would be unstable if not for the fact that we can choose $\pi_\psi=\pi_\Phi\approx 0$ in order to preserve the gauge fixing of the metric, Equations \eqref{eq: approxMetric}. In the Di\'osi model, the kinetic energy term is instead taken to be positive, and its inclusion results in dynamics in $\Phi$ which is continuous, while here, the dynamics in $\Phi$ is discontinuous, owing to the fact that we take the $c\rightarrow \infty$ limit.

Another approach to deriving consistent classical-quantum theories is the measurement 
and feedback approaches of \cite{DiosiHalliwel, tilloy2016sourcing, Kafri_2014}. In these approaches, the classical degree of freedom is sourced by the outcomes of continuous measurements and by construction, such approaches are completely positive and lead to consistent coupling between classical and quantum degrees of freedom. As such, the dynamics for the density matrix of \cite{DiosiHalliwel, tilloy2016sourcing, Kafri_2014} undergoes a stochastic master equation evolution of the general form similar to the unravelling of the quantum state given in Equation \eqref{eq: constraintfinal}. In the special case where $D_0(\Phi_t;x,y)$ and $\sigma(\Phi_t;x,y)$ do not depend on $\Phi_t$, our Equation \eqref{eq: constraintfinal_quantum} can be put into the form of Equation (24) of \cite{tilloy2016sourcing}. When we impose the $\pi_\Phi\approx 0$ constraint and turn the Newtonian field into white noise, we pick up an extra decoherence term in the It\^o formalism, as they do, which is necessary for the normalisation of the quantum state. 

In \cite{tilloy2016sourcing, Kafri_2014}, the noise instead emerges because the Newtonian potential is sourced by the outcome of a continuous measurement of the mass operator. The behaviour in these models is qualitatively the same as those presented here when $D_0(\Phi_t;x,y)$ and $\sigma(\Phi_t;x,y)$ do not depend on $\Phi_t$, meaning that the Newtonian potential diffuses by an amount that depends on the inverse of the strength of the measurement, whilst the quantum system decoheres into its mass eigenbasis because it's being continuously measured. Another difference is that these works generally utilized the mass density operator of a localised particle in the position basis (smeared by a Gaussian in Tilloy-Di\'osi), such that the decoherence of the quantum system emerges from mass measurements in the position basis of a point particle. The measurements are in general imagined to be undertaken using an entangled measuring device, in order to obtain the correlations required to obtain the Di\'osi-Penrose kernel. Here, matter is treated as a quantum field in the non-relativistic limit, although one could consider the point-particle limit by writing the mass density operator as a sum of delta functions in the position basis and integrating over the $\mathcal{D}x^\pm$ position branches in the path integral.

Furthermore, \cite{Kafri_2014, tilloy2016sourcing} imagine the results of a weak external measurement or collapse model as sourcing the gravitational field, and thus the Newtonian potential changes discontinuously, since the results of each measurement can be different. Here we emphasise that it is merely the coupling of quantum matter to the classical gravitational field which is responsible for the localisation of particles. The local time coordinate and the shift $n_i$ are changing stochastically, as can be seen via Equations \eqref{eq: newtUnrav}-\eqref{eq:statechange}, in order to maintain the primary constraint $\pi_\Phi\approx 0$, while the quantum state and Newtonian potential stochastically change in lock-step together.
No measurement postulate nor Born rule is needed, and 
there is no need to think about the ad-hoc field introduced in spontaneous collapse models \cite{grw85, Pearle:1988uh, PhysRevA.42.78, BassiCollapse, PhysRevD.34.470}. Instead, the fully classical treatment of the gravitational degrees of freedom acts to {\it classicalise} the quantum system. 
Although it appears as if the state of the matter fields undergo decoherence, 
there is no decoherence if we condition on the gravitational field. The quantum state is pure conditioned on the classical trajectory when the decoherence vs diffusion trade-off is saturated \cite{UCLhealing,UCLcoherence}. It is only when the gravitational field is integrated out that there appears to be loss of quantum information. 

Although the theory considered here is not predicated on it explaining measurement or collapse, it may still suffer from anomalous heating of the quantum system \cite{bps,ghirardi1986unified,ballentine1991failure,pearle1999csl,bassi2005energy,adler2007lower,lochan2012constraining,nimmrichter2014optomechanical,bahrami2014testing,laloe2014heating,bahrami2014proposal,goldwater2016testing,tilloy2019neutron,donadi2020underground} which constrain collapse models. Since the decoherence couplings $D_0$ can be made arbitrarily small here, albeit at the expense of a large amount of diffusion, it is unclear the extent to which heating bounds constrain the theory, and more investigation is needed here. While the results of \cite{tilloy2017principle}, suggest that the heating might be significant, there is evidence that relativistic effects need to be taken into account \cite{UCLDMDNE}.


Given the novelty of our approach, we used a number of different methods to check the validity of our results. Firstly, we used both the path integral and master equation pictures of the CQ framework to arrive at the main results, namely the behaviour of the classical Newtonian gravitational field coupled with quantum matter. Secondly, the derivation of the Newtonian limit of the CQ framework is done both through a configuration space approach from the full covariant general relativistic theory in Section~\ref{sec: fulltheoryNL} and from a reduced phase space picture in both Section~\ref{sec: CQnewtlim} and in Appendix~\ref{app: PInewt}. Moreover, the same Newtonian limit has also been derived from a scalar theory of gravity which is diffeomorphism invariant~\cite{UCLNordstrom}.

Although the primary motivation for studying the weak field limit of \cite{UCLpost_quantum,UCLPISHORT} is to derive experimental bounds, there are also a number of lessons that can be learned, not only for classical-quantum theories but also attempts to quantise gravity. For example,
we would like to point the reader to an analogy between the process of quantizing a theory and that of adding stochastic noise/diffusion to it when the theory has first-class constraints. When attempting to quantize a theory with constraints, it is well known that two of the possible approaches are Dirac quantization and reduced Phase Space quantization. The former consists in first constructing a kinematical Hilbert space where the classical phase space functions that have been elevated to operators can act on the quantum states and then imposing the constraints quantum mechanically as operator conditions to distinguish physical states. In other words, physical states are zero eigenvalues states of the constraint operator. On the other hand, reduced phase space quantization first factorizes the constraint surface with respect to the action of the gauge group generated by the constraints. This serves to identify the physical degrees of freedom directly at the classical level. Then, the resulting Hamiltonian system is quantized as a usual unconstrained system. The two procedures are not always equivalent, and the relationship between the approaches is discussed at length in the literature
\cite{Kuchar:1986ji,Kuchar:1986jj,Romano:1989zb,Schleich:1990gd, Loll:1990rx,Ashtekar:1991hf}.

In the same way, we could insert a noise process in a Hamiltonian system before or after having reduced the phase space with respect to its constraints. In the main body, we start from the full CQ theory of general relativity, where noise is present in the metric, and reduce it to the Newtonian limit by implementing the constraints. On the other hand, in Appendix \ref{app: PInewt}, we chose the latter approach, restricting the classical degrees of freedom before inserting them in the CQ framework which implements a noise process. It is perhaps remarkable that the two procedures give the same theory here, while in the quantum case, they generally do not.
In this work, we have arrived at this behaviour in complete generality
from a reduction of the CQ degrees of freedom of the relativistic theory,
with the diffusion of the Newtonian potential and decoherence on the quantum system described by the parameters $D_2(\Phi;x,y), D_0(\Phi;x,y)$ satisfying the decoherence/diffusion trade-off. The weak field CQ theories we studied gave a generic prediction: the Newtonian potential diffuses away from its average solution, and in order for the dynamics to be completely positive, the amount of the diffusion is lower bounded by the coherence time for masses in superposition. This is most elegantly described via the path integral formulation of Equation~\eqref{eq: PQG-actionNewton2}.

There are a number of proposals to test the quantum nature of gravity via gravitationally induced entanglement or coherence that are expected to be realizable within the next few decades with technological advancements  \cite{kafri2013noise,kafri2015bounds,bose2017spin,PhysRevLett.119.240402, Marshman_2020,pedernales2021enhancing,carney2021testing,christodoulou2022locally,Danielson:2021egj,lami2023testing}. The idea is that if the underlying theory is local, then witnessing entanglement would imply that gravity is not a classical field. Within the framework of consistent classical-quantum coupling, we are able to inquire from the other direction, asking about the general experimental signatures of treating the gravitational field as being classical.

\begin{table}

\bgroup
\def\arraystretch{2.9}
\begin{tabular}{@{\extracolsep{\fill}} |c| l| l|}\hline  \label{table: kernels}

Master Equation & 
 Decoherence & Diffusion \\  
 \hline 
Continuous (local) & \makecell[l]{$D_0(\Phi; x,y) =D_0(\Phi)\delta(x,y)$} & $D_2(\Phi; x,y) =D_2(\Phi)\delta(x,y)$ \\ 
\hline
Continous (Gaussian) &  \makecell[l]{$D_0^{\alpha\beta}(x,y)=\frac{\lambda^{\alpha\beta}}{m_0^2}g_{\mathcal{N}(x,y)}$} &   $D_2(x,y)=\frac{1}{8}\frac{m_0^2}{r_0^3\lambda}F(x,y)g_{\mathcal{N}(x,y)}$ \\ 
    \hline
Continuous (D.P)  & 
\makecell[l]{$D_0^{\alpha\beta}(x,y)=\frac{D_0^{\alpha\beta}}{|x-y|}$} & $D_2(x,y)=\frac{1}{8}\frac{(D_0^{-1})}{4\pi}\nabla^2_y(\delta(x,y))$  \\ 
 \hline
\end{tabular}
\caption{Possible choices of kernels for the continuous master equations and the resulting diffusion/decoherence coefficients, which are assumed to saturate the trade-off in Equation \eqref{eq: decdiffcont}. For a more detailed study of these kernels, including calculations of the diffusion and decoherence they produce, we refer the reader to \cite{UCLdec_Vs_diff}.}
\egroup
\end{table}

If the Lindbladian coupling in Equation \eqref{eq: contbackreaction} $D_0(\Phi;x,y)$ is ultralocal, the dynamics do not generate entanglement between spatially separated regions, meaning that the models with local couplings parameterize the general form of the continuous master equation which would be ruled out by entanglement witnesses in GIE experiments. 
We give three examples of kernels $D_0, D_2$ for continuous master equations in Table \ref{table: kernels}. 
Models with ultralocal couplings form perhaps the most natural class of CQ dynamics. Non-relativistic versions of these models have already been ruled out by considerations of the decoherence diffusion trade-off \cite{UCLdec_Vs_diff}.  In other words, (in line with assumptions 1-4) classical-quantum Newtonian theories of gravity, which have continuous gravitational degrees of freedom with local interactions and correlations, are already ruled out by experiment. However, relativistic effects play an important role and need to be taken into account \cite{UCLDMDNE}.

We have here considered the case where the gravitational field is considered to be fundamentally classical. 
As an effective theory of Newtonian gravity, we still expect the path integrals and the unravellings derived in this paper to be valid dynamics with a time-local description \cite{UCLpost_quantum}. However, in general, one expects an effective theory to be non-Markovian in some regimes, which means that the couplings $D_0, D_2$ need not be positive semi-definite for all times, \cite{Hall2014, Breuer2016}, nor satisfy the decoherence-diffusion trade-off for all times since this is a consequence of the Markovian assumption. 


\section*{Acknowledgements}
We would like to thank Antoine Tilloy for his insightful comments on an earlier draft of this work. We would like to thank Maite Arcos, Lajos Di\'osi, Gary Horowitz, Jorma Louko, Don Marolf, Emanuele Panella, Andy Svesko, and Bill Unruh for valuable discussions.
J.O. is supported by an EPSRC Established Career Fellowship, and a Royal Society Wolfson Merit Award, A.R. acknowledges financial support from EPSRC and UCL Physics Department, I.L. acknowledges financial support from EPSRC. This research was supported by the National Science Foundation under Grant No. NSF PHY11-25915 and by the Simons Foundation {\it It from Qubit} Network.  Research at Perimeter Institute is supported in part by the Government of Canada through the Department of Innovation, Science and Economic Development Canada and by the Province of Ontario through the Ministry of Economic Development, Job Creation and Trade.

\appendix


\section{Weak field and Newtonian limit of Einstein's equation}\label{app: NewtLimitFromGR}
In this appendix we recall the Newtonian limit of Einstein's equations \cite{CarrollNotes}. To begin, we perform a scalar-vector-tensor decomposition of the metric
 \begin{equation}
        ds^2=-c^2\left(1+\frac{2\Phi}{c^2}\right)dt^2+ \frac{w_i}{c}\big(dt dx^i+dx^i dt\big)+\left[\left(1-\frac{2\psi}{c^2}\right)\delta_{ij}+\frac{2s_{ij}}{c^2}\right] dx^i dx^j,
    \end{equation}
where $s_{ij}$ is traceless and the factors of $c$ ensure that the fields $\Phi, \psi, w_i,s_{ij}$ all have dimensions $c^2$. To arrive at the Newtonian limit, we choose the transverse gauge, which amounts to taking a gauge such that $\partial_i w^i=0$, $\partial_i s^{ij}=0$. We shall also assume we take the rest frame of a particle with mass density $m(x)$, so that $T_{00} = c^4 m(x)$.
Before the gauge choice, the Einstein equation's $G_{\mu \nu}=\frac{8 \pi G}{c^4} T_{\mu \nu}$ look like:
    \begin{align}
        &G_{00}=2\nabla^2\psi=8\pi Gm(x)\\
        &G_{0i}=\frac{2}{c^3}\partial_0\partial_i\psi-\frac{1}{2 c^2}\nabla^2w_i=0\\
        &G_{ij}=\frac{1}{c^2}(\delta_{ij}\nabla^2-\partial_i\partial_j)(\Phi-\psi)-\frac{1}{c^3}\partial_0\partial_{(i}w_{j)}+\frac{2}{c^2}\delta_{ij}\partial_0\partial^0\psi-\frac{1}{c^2}\square s_{ij}=0.
    \end{align}
While after, they reduce to:
    \begin{align}
        &G_{00}=\nabla^2\psi=4\pi Gm(x)\\
        &G_{0i}=\frac{2}{c^3}\partial_0\partial_i\psi - \frac{1}{2 c^2}\nabla^2w_i=0\\
        &G_{ij}=\frac{1}{c^2}(\delta_{ij}\nabla^2-\partial_i\partial_j)(\Phi-\psi)- \frac{2}{c^4}\partial_0 \partial_0 \psi + \frac{1}{c^4}\partial_0 \partial_0 s_{ij} - \frac{1}{c^2}\nabla^2 s_{ij}=0.
    \end{align}
We remind the reader that the $G_{00}$ and $G_{0i}$ components are first order in time derivatives and hence are the constraints on the initial data of the theory, whilst $G_{ij}$ describes the dynamics.  

With this in mind, let us arrive at the Newtonian limit. We first solve the $G_{00}$ component of the Einstein equation, which is the Poisson equation for $\psi$. We see from $G_{0i}$ and the solution for $G_{00}$ that $\partial_0 \partial_i \psi=0$, which imposes that there can be no vector perturbations $w_i =0$. Conversely, we see from the momentum constraint $G_{0i}$ that if there are no vector perturbations, $w_i=0$ then the constraint $\partial_i \partial_0 \psi =0$  must be imposed.

To obtain the final form of the Newtonian limit. We take the trace of $G_{ij}$ to see which imposes that $\psi = \Phi =const$,  which in combination with the fact that $\partial_0 \partial_i \psi=0$ imposes $s_{ij}=0$. 

Altogether, we are then left with the Newtonian metric 
 \begin{equation}
        ds^2=-c^2\left(1+\frac{2\Phi}{c^2}\right)dt^2+\left(1-\frac{2 \Phi}{c^2}\right)\delta_{ij} dx^i dx^j,
    \end{equation}
where $\Phi$ solves the Poisson equation due to  the $G_{00}$ component of Einstein's equation
\begin{equation}
    \nabla^2\Phi=4\pi Gm(x).
\end{equation}


\section{Equivalence of the weak field path integral and master equation}
\label{app: PInewt}

In this appendix, we arrive at the path integral \eqref{eq: PathNewt} from the master equation \eqref{eq: contbackreaction}. To arrive at the path integral, we will use the result of \cite{UCLPILONG}, which derived the correspondence between CQ master equations and path integrals. This derivation shows that the two approaches given in the main body, i.e. formulating the CQ weak field limit by either gauge fixing the full path integral or constructing the master equation from the reduced Hamiltonian, are equivalent. 
We refer the reader directly to \cite{UCLPILONG} for more details on deriving CQ path integrals. 

We start by recalling that because the equation of motion of the weak field total Hamiltonian \eqref{eq: totham}:
\begin{equation}
     \dot{\psi} = - \frac{4G \pi c^2 \pi_{\psi}}{3} - \frac{1}{12} \partial_i n^i, \ \dot{\pi}_{\psi} = \frac{\nabla^2( \Phi - \psi)}{4\pi G}, \ \dot{\Phi} = \lambda_{\Phi}, \  \dot{\pi}_{\Phi}= \frac{\nabla^2 \Phi}{4 \pi G } -m , \ \dot{n}^i = \lambda^i, \  \dot{\pi}_i = -\frac{1}{12} \partial_i \pi_{\psi}
\end{equation}
only associate back-reaction to $\pi_{\Phi}$, the path integral takes the form:

\begin{equation}
\label{eq: PIHamiltonianFull}
    \begin{split}
  \varrho(z_f,m^+_f,m^-_f,t_f)& = \mathcal{N}\int \mathcal{D}z \mathcal{D}m^+\mathcal{D}m^- \delta\left( \dot{\psi}  + \frac{4G \pi c^2 \pi_{\psi}}{3} + \frac{1}{12} \partial_i n^i\right) \delta\left( \dot{\pi}_{\psi} - \frac{\nabla^2( \Phi - \psi)}{4\pi G}\right)\delta(\dot{\Phi} - \lambda_{\Phi}) \\&\quad\quad \times \delta(\dot{n}^i-\lambda^i) \delta \left( \dot{\pi_i} +\frac{1}{12}\partial_i \psi\right) \delta(\pi_{\Phi})e^{\mathcal{I}_{CQ}[z,m^+,m^-,t_i,t_f]}\varrho(z_i,m^+_i,m^-_i,t_i)
  \end{split}
\end{equation}
where the last delta function imposes the Newtonian limit constraint $\pi_{\Phi} \approx 0$. In Equation \eqref{eq: PIHamiltonianFull}, for the sake of clarity, we have summarised all the classical degrees of freedom with $z$ such that the functional measure over the classical functions $\mathcal{D} z$ represents: 
\begin{equation}
    \mathcal{D}z  = \mathcal{D}\psi\, \mathcal{D}\pi_{\psi} \,
    \mathcal{D}\vec{n} \, \mathcal{D}\vec{\pi}\,
    \mathcal{D}\lambda_{\Phi}\,
    \mathcal{D}\vec{\lambda}\,
     \mathcal{D}\Phi\,
  \mathcal{D}\pi_{\Phi}
\end{equation}
and the hybrid action $\mathcal{I}_{CQ}$ is:
\begin{equation}
\label{eq: CQNewtActionApp}
    \begin{split}
    \mathcal{I}_{CQ}[z, m^+,m^-,t_i,t_f] &=  \int_{t_i}^{t_f} dtd\vec{x}\,\bigg[ i\big(\mathcal{L}_{\mathcal{Q}}[m^+] - \mathcal{V}_I[\Phi,m^+] - \mathcal{L_{\mathcal{Q}}}[m^-] +\mathcal{V}_I[\Phi,m^-] \big) \\
    & \quad  -\frac{D_0[z]}{2}\big(m^+(x)-m^-(x)\big)^2 - \frac{1}{2D_2[z]}  \left(\dot{\pi}_{\Phi} -\frac{\nabla^2 \Phi}{4\pi G } + \frac{1}{2}\big(m^+(x) +m^-(x)\big)\right)^2\bigg].
  \end{split}
\end{equation}
where $\mathcal{L}_Q[m^\pm]$ is the matter Lagrangian and $\mathcal{V}_I[\Phi,m^\pm]=  \Phi(x)m^\pm(x) $. The CQ interaction term in Equation \eqref{eq: CQNewtActionApp} is the path integral version of Equation \eqref{eq: contbackreaction}. This correspondence was derived explicitly in \cite{UCLPILONG} and takes the same form as the Hamiltonian CQ path integrals in \cite{UCLPISHORT, UCLPILONG}.

Just as in the deterministic case, one can then reduce the system to describe it in terms of the Newtonian potential alone. Performing all of the delta integrals in Equation \eqref{eq: PIHamiltonianFull}, we arrive at the hybrid Newtonian CQ path integral in terms of $\Phi$ alone, 

\begin{equation}
\label{eq: PathNewtApp}
\varrho(\Phi_f,m^+_f,m^-_f,t_f)= \mathcal{N}\int \mathcal{D} \Phi\mathcal{D}m^+\mathcal{D}m^-  \; e^{\mathcal{I}_{CQ}[\Phi,m^\pm,t_i,t_f]}  \varrho(\Phi_i,m^+_i,m^-_i,t_i), 
\end{equation}
where the CQ action is given by:
\begin{equation}
\label{eq: PathNewtAppAction}
\begin{split}
  \mathcal{I}_{CQ}[\Phi,m^+,m^-,t_i,t_f] & =  \int_{t_i}^{t_f} dtd\vec{x} \,\bigg[i\big(\mathcal{L}_{\mathcal{Q}}[m^+] -\mathcal{V}_I[\Phi,m^+] - \mathcal{L_{\mathcal{Q}}}[m^-] + \mathcal{V}_I[\Phi,m^-] \big) \\
  & - \frac{D_0[\Phi]}{2}\big(m^+(x) - m^-(x)\big)^2 - \frac{1}{2 D_2[\Phi]}\left(\frac{\nabla^2 \Phi}{4 \pi G} -  \frac{1}{2}\big( m^+(x) + m^-(x)\big)\right)^2\bigg]
  \end{split}
\end{equation} 
which takes the same form as the one derived from general relativity in Equation~\eqref{eq: PathNewt}. The apparent difference in form of this path integral with the unravelling in Equations \eqref{eq: constraintfinal} and \eqref{eq: constraintfinal_quantum} arises due to the fact that the path integral in \eqref{eq: PathNewtApp} is unnormalised; one may normalise this path-integral by computing a Gaussian integral over $\nabla^2 \Phi$ and upon doing so one finds the appearance of secondary decoherence and $\hat{V}_m$ terms.





\section{Unravelling of CQ theories}
\label{app: unravelling}
In this appendix, we present some useful details on the topic of the CQ unravelling.

Given a CQ state $\varrho(z,t)=\mathds{E}[\delta(z-Z_t)\rho_t]$, the dynamics which generate the stochastic trajectories of the classical $\{Z_t\}_{t>0}$ and quantum $\{\rho_t\}_{t>0}$ degrees of freedom will induce the stochastic trajectory of the hybrid state $\varrho(z,t)$. The dynamics is positive and norm-preserving and can be written as a series of stochastic differential equations \cite{UCLhealing}

    \begin{align}
    \label{eq: unravel_general}
        &dZ_{t,i}=D_{1,i}(Z_t)dt+\langle D_{1,i}^{\alpha 0}(Z_t)L_\alpha+D_{1,i}^{\alpha 0}(Z_t)L_\alpha^\dagger \rangle dt +\sigma_{ij}(Z_t)dW_j, \\
        &d\rho_t=-i[H(Z_t),\rho_t]dt+D_0^{\alpha\beta}(Z_t)L_\alpha\rho L_\beta^\dagger dt-\frac{1}{2}D_0^{\alpha\beta}(Z_t)\{L_\beta^\dagger L_\alpha,\rho_t\}_+ dt \nonumber\\
        &\quad\quad\quad + D_{1,j}^{\alpha 0}\sigma_{ij}^{-1}(Z_t)(L_\alpha-\langle L_\alpha\rangle)\rho_t dW_i +D_{1,j}^{\alpha 0}\sigma_{ij}^{-1}\rho_t(L_\alpha^\dagger-\langle L_\alpha^\dagger \rangle)(Z_t)dW_i,
    \end{align}
where $dW_i$ is the standard multivariate Wiener process and $\sigma_{ij}$ is defined by $D_{2,ij}^{00}=\sigma_{ik}\sigma_{kj}^T$.

The first equation represents the path of the classical degrees of freedom $Z_i$ through phase space. The first term is the usual classical evolution, and the second describes the back-reaction of the quantum degrees of freedom, appearing through the presence of the Lindblad operators. In contrast, the last term represents the random kicks that cause diffusion. The second equation allows one to simulate paths of the quantum state through Hilbert space. We can distinguish the standard unitary evolution, the decoherence terms (analogous to the GKSL equation but with a dependence on the classical phase space), and the noise in its trajectory manifested in the last two terms. One can notice how these coupled differential equations are much easier to simulate on a computer, and their averaged-out paths will recover the master equation formulation.

Moreover, in \cite{blanchard1995event,UCLunrav, UCLhealing}, it has been argued that the unravelling picture of the CQ state has an added ontological value with respect to the GKSL unravelling. The unravelling of the GKSL is highly non-unique due to the possibility of decomposing the same dynamics using different Lindblad operators. Instead, when unravelling the CQ master equation, if the assumption is made that each Lindblad operator has a one-to-one relation to a shift in the classical degrees of freedom, the resulting unravelling of the dynamics will be unique when conditioned on the classical degrees of freedom. The consequence of this is that stochastic trajectories of the state can be associated with real physical trajectories, and the resulting collapse into a particular state is actually happening due to the physical interaction between the classical and quantum degrees of freedom. In other words, the unravelling allows us to determine the evolution of the quantum state conditioned on the classical trajectory, which remains pure if the decoherence diffusion trade-off is saturated \cite{UCLhealing}.

We now show how to recover the continuous backreaction term of Equation~\eqref{eq: contbackreaction} from the unravelled equation of Equation~\eqref{eq: newtUnrav}. Given the stochastic nature of $\partial_i n^i$, it is not possible to recover a closed form, but we can see how the correlation terms emerge due to the divergence of the shift being a white noise process. 

We start by defining the CQ state as:
    \begin{equation}
        \varrho(\Phi,\pi_\Phi.t)=\mathbb{E}[\delta(\Phi_t-\Phi)\delta(\pi_{\Phi\,t}-\pi_\Phi)\rho_t].
    \end{equation}
When we now take the total differential of the CQ state, we have to apply It\^o's rule:
    \begin{equation}
        \begin{split}
         d\varrho=\mathbb{E}&[d\delta(\Phi_t-\Phi)\delta(\pi_{\Phi\,t}-\pi_\Phi)\rho_t+\delta(\Phi_t-\Phi)d\delta(\pi_{\Phi\,t}-\pi_\Phi)\rho_t\\
         &+\delta(\Phi_t-\Phi)\delta(\pi_{\Phi\,t}-\pi_\Phi)d\rho_t+d\delta(\Phi_t-\Phi)d\delta(\pi_{\Phi\,t}-\pi_\Phi)\rho_t\\
         &+d\delta(\Phi_t-\Phi)\delta(\pi_{\Phi\,t}-\pi_\Phi)d\rho_t+\delta(\Phi_t-\Phi)d\delta(\pi_{\Phi\,t}-\pi_\Phi)d\rho_t+\dotsb],   
        \end{split}        
    \end{equation}
where higher terms of order $\mathcal{O}(dt^2)$ or higher are immediately discarded.

We will now start to unpack the terms one at a time. Recalling that $\Phi$ and $\pi_\Phi$ are functionals, we have to pay attention to how their total derivatives are expanded. We keep only terms of order less than $\mathcal{O}(dt^2)$. usually, this is enough to guarantee a closed form for a continuous master equation. Unfortunately, the fact that $\partial_i n^i \approx \frac{dW_t}{dt}$ means that any power of these terms will never be greater than $\mathcal{O}(dt^2)$. Therefore, we will represent all these terms and their product with terms of order $\mathcal{O}(dt)$ as dots $\dotsb$. We write explicitly only the terms that lead to the continuous part of the master equation and the continuous backreaction.
    \begin{equation}
        \begin{split}
            d\delta(\Phi_t-\Phi)&=\int d^3z\,\frac{\delta}{\delta\Phi_t(z)}\delta(\Phi_t(x)-\Phi(x)) d\Phi_t(z) \\
            &+ \int d^3z d^3w\,\frac{\delta^2}{\delta\Phi_t(z)\delta\Phi_t(w)}\delta(\Phi_t(x)-\Phi(x))d\Phi_t(z)d\Phi_t(w)+\dotsb\\
            &=\int d^3z\,\frac{\delta}{\delta\Phi_t(z)}\delta(\Phi_t-\Phi)\left(-\frac{1}{12}\partial_i n^i\right)dt+\dotsb\\
            &=\frac{1}{12}\int d^3z\,\frac{\delta}+{\delta\Phi(z)}\delta(\Phi_t-\Phi) \partial_i n^i dt+\dotsb,
        \end{split}
    \end{equation}
where in the last line we have used the property of delta functions stating that $\delta_{\Phi_t}\delta(\Phi_t-\Phi)=-\delta_{\Phi}\delta(\Phi_t-\Phi)$ to change the functional derivative variable. Therefore, the first term reduces to:
    \begin{equation}
        \mathbb{E}[d\delta(\Phi_t-\Phi)\delta(\pi_{\Phi\,t}-\pi_\Phi)\rho_t ]=\frac{1}{12}\int d^3x\,\frac{\delta\varrho}{\delta\Phi(x)} \partial_i n^i dt+\dotsb,
    \end{equation}
where we have used the definition of the CQ state.

Proceeding, we have:
    \begin{equation}
        \begin{split}
            d\delta(\pi_{\Phi\,t}-\pi_\Phi)&=\int d^3z\,\frac{\delta}{\delta\pi_{\Phi\,t}(z)}\delta(\pi_{\Phi\,t}(x)-\pi_\Phi(x))d\pi_{\Phi\,t}(z)\\
            &+ \int d^3z d^3w\,\frac{\delta^2}{\delta\pi_{\Phi\,t}(z)\delta\pi_{\Phi\,t}(w)}\delta(\pi_{\Phi\,t}(x)-\pi_\Phi(x))d\pi_{\Phi\,t}(z)d\pi_{\Phi\,t}(w)\\
            &=-\int d^3z\,\frac{\delta}{\delta\pi_\Phi(z)}\delta(\pi_{\Phi\,t}-\pi_\Phi)\left(\frac{\nabla^2\Phi}{4\pi G}-\langle \hat{m}\rangle\right)dt\\
            &-\int d^3z d^3y\,\frac{\delta}{\delta\pi_\Phi(z)}\delta(\pi_{\Phi\,t}-\pi_\Phi)\sigma(\Phi,x,y)dW_t(y)\\
            &+ \int d^3z d^3w d^3y d^3y'\,\frac{\delta^2\delta(\pi_{\Phi\,t}-\pi_\Phi)}{\delta\pi_\Phi(z)\delta\pi_\Phi(w)}\sigma(\Phi,z,y)\sigma(\Phi,w,y')dW(y)dW(y').        
        \end{split}
    \end{equation}
When we now average over the noise, we use the properties of the Wiener process \eqref{eq: Wiener} and the definition of the diffusion coefficient \eqref{eq: diffcoeff}
to arrive at:
    \begin{equation}
    \begin{split}
        \mathbb{E}[\delta(\Phi_t-\Phi)\,d\delta(\pi_{\Phi\,t}-\pi_\Phi)\rho_t ]&=-\int d^3x\,\frac{\delta\varrho}{\delta\pi_\Phi(x)} \left(\frac{\nabla^2\Phi}{4\pi G}-\langle \hat{m}\rangle\right)dt\\
        &\quad +\int d^3x d^3y\,\frac{\delta^2}{\delta\pi_\Phi(x)\delta\pi_\Phi(y)}(D_2(\Phi,x,y)\varrho)dt.
    \end{split}
    \end{equation}
Moving on to the next term, we find 
    \begin{equation}
    \begin{split}
        d\rho_t &=\frac{\partial\rho_t}{\partial t}dt+\frac{\partial\rho_t}{\partial W_t}dW_t+\frac{1}{2}\frac{\partial^2\rho}{\partial W_t^2}dW_t^2\\
        &\quad -i[H_m,\rho_t] dt + \frac{1}{2}\int d^3x\,d^3 y\, D_0(\Phi_t; x,y) [ \hat{m}(x), [ \rho_t, \hat{m}(y) ] ]  dt \\
        & \quad+ \frac{1}{2}\int d^3x\, d^3y\, \sigma^{-1}(\Phi_t;x,y) \big(\hat{m}(x)\rho_t+\rho_t\, \hat{m}(x)-2\rho_t \langle \hat{m}(x) \rangle\big) d W(y),
    \end{split}
    \end{equation}
which gives:
    \begin{equation}
        \mathbb{E}[\delta(\Phi_t-\Phi)\delta(\pi_{\Phi\,t}-\pi_\Phi)\,d\rho_t ]=-i[H_m,\varrho] dt + \frac{1}{2}\int d^3x\, d^3 y\, D_0(\Phi; x,y) [ \hat{m}(x), [ \varrho, \hat{m}(y) ] ]  dt.
    \end{equation}
At this point, we need to consider the expectations of terms with mixed derivatives. After a closer inspection we notice that keeping in mind that we will be averaging over the noise, only one term is relevant for the continuous part of the master equation, specifically:
    \begin{equation}
        \begin{split}
            d\delta(\pi_{\Phi\,t}-\pi_\Phi)\,d\rho_t=-\frac{1}{2}\int d^3x\,d^3z d^3y d^3w &\frac{\delta}{\delta\pi_{\Phi\,t}(z)}\delta(\pi_{\Phi\,t}-\pi_\Phi)\sigma(\Phi,x,y)\sigma^{-1}(\Phi,x,w)\\
            & \times\big(\hat{m}(x)\rho+\rho\, \hat{m}(x)-2\rho \langle \hat{m}(x) \rangle\big) d W(y) dW(w),
        \end{split}
    \end{equation}
while the other surviving mixed terms will include the correlations between the stochastic shift and the classical gravitational field.
When we average over the noise, we can integrate $dw$ over the delta function $\delta(y-w)$ coming from the Wiener processes to use $\int d^3y\, \sigma(\Phi,x,y)\sigma^{-1}(\Phi,x,y)=\mathds{1}$.
Therefore, we arrive at:
    \begin{equation}
        \mathbb{E}[\delta(\Phi_t-\Phi)\,d\delta(\pi_{\Phi\,t}-\pi_\Phi)\,d\rho_t ]=\int d^3x \left(\frac{1}{2}\hat{m}(x)\frac{\delta\varrho}{\delta\pi_{\Phi}(x)}+\frac{1}{2}\frac{\delta\varrho}{\delta\pi_{\Phi}(x)}\,\hat{m}(x)- \frac{\delta\varrho}{\delta\pi_{\Phi}(x)}\langle \hat{m}(x) \rangle\right) dt.
    \end{equation}

Finally, we can sum all the terms and divide by $dt$ to arrive at the Master Equation:
\begin{equation}
    \begin{split}
        \frac{\partial\varrho}{\partial t}&=\frac{1}{12}\int d^3x\,\frac{\delta\varrho}{\delta\Phi(x)} \partial_i n^i -\int d^3x\,\frac{\delta\varrho}{\delta\pi_\Phi(x)} \left(\frac{\nabla^2\Phi}{4\pi G}-\langle \hat{m}\rangle\right)\\
        &\quad -i[H_m,\varrho]  + \frac{1}{2}\int d^3 y\, D_0(\Phi; x,y) [ \hat{m}(x), [ \varrho, \hat{m}(y) ] ]\\
        &\quad +\int d^3x \left(\frac{1}{2}\hat{m}(x)\frac{\delta\varrho}{\delta\pi_{\Phi}(x)}+\frac{1}{2}\frac{\delta\varrho}{\delta\pi_{\Phi}(x)}\,\hat{m}(x)- \frac{\delta\varrho}{\delta\pi_{\Phi}(x)}\langle \hat{m}(x) \rangle\right)\\
        &\quad +\int d^3x d^3y\,\frac{\delta^2}{\delta\pi_\Phi(x)\delta\pi_\Phi(y)}(D_2(\Phi,x,y)\varrho)+\dotsb.
    \end{split}
\end{equation}
We notice how the terms proportional to the expectation value $\langle\hat{m}\rangle$ simplify, and we arrive at:
    \begin{equation}
    \begin{split}
        \frac{\partial\varrho}{\partial t}&=-i[H_m,\varrho]+\frac{1}{12}\int d^3x\,\frac{\delta\varrho}{\delta\Phi(x)} \partial_i n^i -\frac{1}{4\pi G}\int d^3x\,\frac{\delta\varrho}{\delta\pi_\Phi(x)}\nabla^2\Phi\\
        &\quad   + \frac{1}{2}\int d^3 y\, D_0(\Phi; x,y) [ \hat{m}(x), [ \varrho, \hat{m}(y) ] ]+\int d^3x d^3y\,\frac{\delta^2}{\delta\pi_\Phi(x)\delta\pi_\Phi(y)}(D_2(\Phi,x,y)\varrho)\\
        &\quad +\frac{1}{2}\int d^3x \left(\hat{m}(x)\frac{\delta\varrho}{\delta\pi_{\Phi}(x)}+\frac{\delta\varrho}{\delta\pi_{\Phi}(x)}\,\hat{m}(x)\right)+\dotsb,
    \end{split}
\end{equation}
which we can rewrite as:
\begin{equation}
    \begin{split}
    \frac{\partial\varrho}{\partial t}&=-i[H_m,\varrho]+\frac{1}{12}\int d^3x\,\frac{\delta\varrho}{\delta\Phi(x)} \partial_i n^i -\frac{1}{4\pi G}\int d^3x\,\frac{\delta\varrho}{\delta\pi_\Phi(x)}\nabla^2\Phi\\
    &+\frac{1}{2}\int d^3x \big( \{\mathcal{H}^I(x),\cqstate\}-\{\cqstate,\mathcal{H}^I(x)\} \big) +\int d^3x d^3y\, D_{2}(\Phi; x,y)\{\mathcal{H}_C(x),\{\cqstate,\mathcal{H}_C(y)\}\}  \\
    &\quad\quad\quad + \frac{1}{2}\int d^3 x d^3y\, D_0(\Phi;x,y)\left( [ \hat{m}(x), [ \cqstate, \hat{m}(y) ] ] \right) + \dotsb,
    \end{split}
\end{equation}
where $\mathcal{H}_I$ and $\mathcal{H}_C$ are the interaction Hamiltonian density and the Hamiltonian density for the classical degrees of freedom as specified in Section~\ref{sec: Newtlim} and $\dotsb$ includes the jumping terms and the correlation terms present due to the stochastic nature of $\partial_i n^i$.

\bibliography{NewtLimitbib.bib}
\bibliographystyle{apsrev}

\end{document}